\documentclass[conference]{IEEEtran}
\IEEEoverridecommandlockouts
\usepackage{cite}
\usepackage{amsmath,amssymb,amsfonts}
\usepackage{algorithmic}
\usepackage{graphicx}
\usepackage{textcomp}
\usepackage{xcolor}
\usepackage{url}
\usepackage{subcaption,lipsum}
\usepackage{enumitem}
\setlist{nosep}
\usepackage{float}
\usepackage[export]{adjustbox}
\usepackage{hyperref}
\emergencystretch=\maxdimen
\def\BibTeX{{\rm B\kern-.05em{\sc i\kern-.025em b}\kern-.08em
    T\kern-.1667em\lower.7ex\hbox{E}\kern-.125emX}}
\begin{document}
\pagestyle{plain}

\title{Memory Efficient and Staleness Free Pipeline Parallel DNN Training Framework with Improved Convergence Speed}

\author{
\IEEEauthorblockN{Ankita Dutta}
\IEEEauthorblockA{Machine Intelligence Unit \\
Indian Statistical Institute\\
203 Barrackpore Trunk Road, \\Kolkata 700108, India.}
\and
\IEEEauthorblockN{Nabendu Chaki}
\IEEEauthorblockA{Department of Computer Science and Engineering \\
University of Calcutta\\
JD-2, Sector III, Salt Lake City, \\Kolkata 700098, India.}
\and
\IEEEauthorblockN{Rajat K. De\textsuperscript{\textdagger}}
\IEEEauthorblockA{Machine Intelligence Unit \\
Indian Statistical Institute\\
203 Barrackpore Trunk Road, \\Kolkata 700108, India.\\
\textsuperscript{\textdagger}Corresponding author: \textbf{rajat@isical.ac.in}}
}

\maketitle

\begin{abstract}
High resource requirement for Deep Neural Network (DNN) training across multiple GPUs necessitates development of various parallelism techniques. In this paper, we introduce two interconnected DNN training frameworks, namely, V-TiMePReSt and I-TiMePReSt, based on pipeline parallelism, a variant of model parallelism. V-TiMePReSt is a completely staleness-free system which enables the DNNs to be trained on the latest updated weights in each stage of all forward and backward passes. Developing staleness-aware systems at the expense of weight stashing reduces GPU-memory consumption, however, increases the number of epochs to converge. Thus, we introduce I-TiMePReSt, which is also a staleness-aware system, but not at the expense of weight stashing. It does not rely solely on the stale weights or the latest updated weights. I-TiMePReSt computes an intermediate weight towards the latter and performs backward pass on it. Additionally, we formulate the significance of the stale weights mathematically depending on the degree of staleness. In contrast to V-TiMePReSt, I-TiMePReSt works based on the assumption that stale weights have a significant contribution in training, which can be quantified mathematically based on the degree of staleness, although there are other contributory factors which should not be ignored. Experimental results show that V-TiMePReSt is advantageous over existing models in terms of $1)$ the extent of staleness of the weight parameter values and $2)$ GPU memory efficiency, while I-TiMePReSt is superior in terms of $1)$ removing staleness of the weight parameters without removing weight stashing and $2)$ maintaining the trade-off between GPU memory consumption and convergence speed (number of epochs).
\end{abstract}
\vspace{0.5em}

\section{Introduction}
\label{sec:Introduction}
DNNs have shown remarkable efficiency in a diverse range of applications. DNNs can capture complex patterns in high-dimensional data. They have been successfully applied to various fields of research, such as computer vision, natural language processing, and healthcare, where they achieve significant advancements with improved outcomes \cite{10049507}. As their applicability expands, the complexity of these models has increased significantly. Consequently, extensive computational resources are required for training. Advancements in low-cost and high-capacity data acquisition technologies have led to an exponential growth in the volume of available data. Thus, the increasing amount and size of nonlinear data have made it necessary to use deeper and more complex neural network models. Training large-scale DNNs has become computationally expensive, both in terms of time and hardware resources \cite{LIU2024317}. However, with sufficient computational capacity, DNNs can achieve substantial performance improvements if and only if a balanced use of all resources is maintained with no under or over-utilization \cite{kim2023bpipe}.

The primary factor contributing to the rise in computational demands is the growing number of operations required for executing forward and backward passes during training. More complex models, having greater number of layers, neurons, or advanced architectures, inherently involve a higher volume of mathematical computations. Similarly, larger models necessitate a greater number of parameters, each of which must be stored in memory, thereby increasing the overall
memory footprint. Additionally, memory is consumed by the intermediate parameter values generated during computation, the
quantity of which scales with model complexity. Both the computational load and memory footprint associated with training and inference increase proportionally with the complexity of DNNs \cite{chilimbi2014project}.

The rapid and exponentially rising computational demands of DNN training have led to the development of specialized hardware accelerators, such as Google Tensor Processing Units (TPUs), NVIDIA Tesla GPUs, and Xilinx Alveo FPGA-based accelerators, among others \cite{unnikrishnan2021layerpipe}. Despite the efficiency of these high-performance computing solutions, their high cost presents a significant barrier to accessibility. As an alternative, distributed training across multiple commodity hardware systems has gained relevance, which poses as a cost-effective solution compared to individual high-end servers. However, transitioning from conventional single-machine deep learning to distributed training paradigm remains a challenge. Thus, it has become necessary to explore and identify appropriate distributed programming environment in order to develop more cost-effective parallelization strategies to facilitate the efficient deployment of distributed deep learning models. Parallelization strategies can be classified into two categories such as data parallelism and model parallelism. Model parallelism \cite{brakel2024model} acts as an effective strategy to address the issue of substantial computational demands by partitioning the DNN itself across the cluster.  

The partitioning can be performed at different levels, leading to two primary variants, namely tensor parallelism  \cite{shoeybi2019megatron, yi2022optimizing, jiang2024megascale} and pipeline parallelism \cite{narayanan2019pipedream, huang2019gpipe, boral2023anomaly}. In tensor parallelism, individual neurons or computational nodes are distributed across devices, whereas pipeline parallelism involves distributing entire layers of
the network over the devices. Large-scale DNNs often surpass the memory capacity of GPUs. Thus, data parallelism poses as insufficient for their training. Pipeline parallelism is a promising approach for training large DNNs due to ease of implementation and partitioning, and is typically used with data parallelism \cite{kim2023bpipe, zhao2021v}. 

Among various parallelization strategies in deep learning, data parallelism or pattern parallelism \cite{ben2019demystifying, chen2022sapipe, LI2021206} is the most straightforward in terms of implementation. Each machine in a distributed cluster maintains a complete copy of the entire DNN in memory. The model parameters are replicated across multiple processors or computers \cite{raina2009large, jacobs2024system}. The training dataset is partitioned and allocated across those machines, enabling independent execution of forward and backward passes in each machine on its assigned data shard. Before updating the model weights, the computed gradients must be synchronized \cite{shallue2019measuring, ben2019demystifying, li2020pytorch} to obtain the final gradient across the entire training dataset. The Parameter Server Architecture \cite{cui2016geeps, dean2012large, li2013parameter} and the All-Reduce Architecture or Collective Communication \cite{narayanan2019pipedream, paszke2019pytorch} are two different synchronization strategies, both introduced by Baidu Research\footnote{\url{https://github.com/baidu-research/baidu-allreduce}}. Upon completion of these computations, weight updates are performed based on the aggregated gradients from all independent training instances. The gradient synchronization is performed only once in each training iteration \cite{kim2023bpipe}.

In contrast, model parallelism \cite{LI2021206} is inherently more complex than data parallelism in terms of implementation. Instead of distributing the dataset across multiple machines, model parallelism partitions the DNN itself across the cluster of devices. Unlike data parallelism, model parallelism does not inherently require the training dataset to be distributed among machines.

While data parallelism effectively scales to large datasets by leveraging distributed computation, it is constrained by the memory requirements of storing the entire model on each machine. The replication of all tthe model parameters, optimizer states, and gradients across the machines becomes infeasible in case of large DNNs \cite{jacobs2024system, li2023colossal}. Model parallelism, on the other hand, enables the training of extremely large networks that exceed the memory capacity of a single device, thereby making it a promising approach for handling highly complex architecture. Selecting an appropriate parallelization strategy depends on the factors such as model size, dataset volume, and computational constraints.
\begin{figure}[!ht]
  \begin{subfigure}{.25\textwidth}
  \centering
    \includegraphics[width=0.8\linewidth, height = 2.5cm]{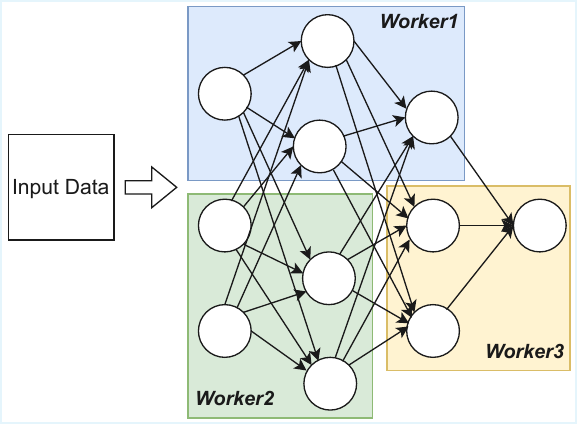}
    \caption{Tensor Parallelism}
    \label{fig:Tensor Parallelism}
  \end{subfigure}
  \begin{subfigure}{.2\textwidth}
  \centering
    \includegraphics[width=0.8\linewidth, height = 2.5cm]{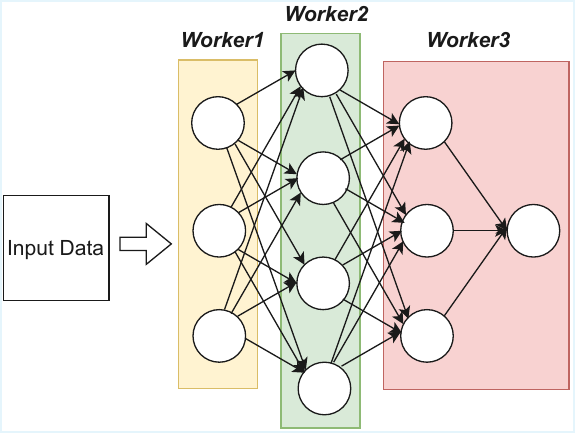}
    \caption{Pipeline Parallelism}
    \label{fig:Pipeline Parallelism}
  \end{subfigure}
  \label{fig:Tensor and Pipeline Parallelism}
  \caption{\textbf{Example tensor and pipeline parallelism mechanisms with three worker nodes. A DNN with four layers is distributed across the worker nodes in two ways. In tensor parallelism, the distribution is based on intra-layer partitions, and in pipeline parallelism, the distribution is layer-wise. Each mini-batch of input data is passed through all the consecutive stages.}}
\end{figure}

In this paper, we have designed two pipeline parallelism frameworks, viz, V-TiMePReSt (TiMePReSt with removed vertical weight stashing) and I-TiMePReSt (TiMePReSt with intermediary weight computation), distributing the network layers across multiple machines (accelerators), working together with data parallelism. Both the models are advanced versions of TiMePReSt \cite{dutta2024timeprest}, and have their own advantages and disadvantages. V-TiMePReSt is advantageous over the existing models in terms of $1)$ the extent of staleness of weight parameters values and $2)$ GPU-memory efficiency. To the best of our knowledge, no models exist in literature which is completely free of staleness, unlike V-TiMePReSt. This property makes it unique among all pipeline parallelism based DNN training frameworks. Besides, the minimal GPU-memory consumption reduces the requirement of high GPU-memory servers. However, the late convergence in terms of the number of epochs is a bottleneck of V-TiMePReSt. 

I-TiMePReSt is another model introduced in this paper, which is advantageous in terms of $1)$ removing staleness of weight parameters without removing weight stashing and $2)$ achieving trade-off between GPU-memory consumption and convergence speed (number of epochs). Experiments demonstrate that discarding stale weights decreases convergence speed. Thus, I-TiMePReSt assumes that both stale and updated weights have significant contributions in the training process of a DNN. We have formulated a mathematical expression to calculate the significance of each of them computationally, which is a highly distinctive approach compared to the existing systems. It also introduces a formalism to calculate a progressive weight value between the stale and actual updated weights based on the computed significances. The newly computed intermediate weights are used in the backward pass during training. Experiments show that the entire approach not only improves convergence speed, it also avoids relying solely on either stale or updated weights, which makes it unique among the existing systems.

\section{Background and Related Works}
\label{sec:Background and Related Works}
This section provides a brief overview of pipeline parallelism used for training DNN as the proposed systems are based on this type of parallelism.

\subsubsection{Pipeline Parallelism}
\label{sec:Pipeline Parallelism}
Pipeline parallelism, also referred to as layer-wise parallelism \cite{AKINTOYE2023432} or inter-layer parallelism in the context of DNN training, enables the sequential decomposition of a DNN into layer-wise partitions, which are then allocated across multiple GPUs. As illustrated in Figure \ref{fig:Pipeline Parallelism}, each partition consists of one or more consecutive layers, with computations distributed accordingly. The output of each partition is subsequently transmitted to the next device in the pipeline for further processing. To design an efficient pipeline-based distributed DNN training framework, several critical challenges must be addressed.
\begin{itemize}
\item \texttt{Weight Consistency:} During training, mini-batches at different stages of the pipeline may still be executing while weight parameters are being updated. A key question is determining which version of the model weights should be used for subsequent mini-batches and what level of weight staleness is permissible \cite{chen2016revisiting} in such a scenario.
\item \texttt{Performance Bottlenecks:} Pipeline parallelism is often associated with time inefficiencies, which can hinder training speed. Effective strategies are required to mitigate bottlenecks and enhance computational efficiency.
\item \texttt{GPU-Memory Utilization and offloading:} Given the significant memory overhead associated with pipeline-based training, it is crucial to develop methods for optimizing GPU-memory usage to ensure efficient resource allocation and scalability. While it is implemented in practice, it is also important to perform memory offloading \cite{wan2025pipeoffload} while maintaining a trade-off between memory and pipeline throughput. 
\end{itemize}

In data parallelism, weight versioning is not performed as the weight parameters are updated only once at the end of each epoch. However, in distributed DNN training frameworks based on model parallelism, particularly pipeline parallelism, weight versioning within an epoch plays a crucial role in ensuring convergence. Unlike data parallelism, pipeline parallelism involves multiple weight updates within a single epoch, typically occurring after the completion of each mini-batch. The frequency of update and the update strategy are critical research challenges that directly impact training efficiency. A key issue in pipeline parallelism is the staleness of weights across mini-batches, arising due to multiple updates within an epoch. The degree of staleness of the weights significantly affects both the convergence of the model and performance optimization. 

Pipeline parallelism performs less communication compared to data parallelism \cite{narayanan2019pipedream}. Data parallelism has to aggregate gradients for all parameters from all the workers and send the updated weights to them (either using collective communication or a parameter server). In contrast, pipeline parallelism communicates the gradients and the output activations for a few layers (allocated to a single machine) to only a single other worker. This one-to-one uni-directional communication may result in large reductions in communication \cite{narayanan2019pipedream}.

Two commonly used pipeline parallelism frameworks for distributed large-scale deep learning are PipeDream \cite{narayanan2019pipedream} and GPipe \cite{huang2019gpipe}. In order to achieve improved pipeline throughput, PipeDream integrates intra-batch and inter-batch parallelism. The integration enables the overlap of computation and communication, as well as reduced communication. It retains the same weight versions throughout the forward and backward passes of a mini-batch, as well as all the stages in a pass, leveraging both vertical and horizontal weight stashing. The need to maintain weight consistency increases the memory footprint for each stage of DNN training on a mini-batch. 

A backward pass begins as soon as a forward pass concludes, using the same set of workers in reverse order. This bidirectional training of DNNs is devised in PipeDream. PipeDream introduces a scheduling strategy, named as $1$F$1$B ($1$ Forward $1$ Backward) scheduling, which is capable of avoiding circular waiting between forward and backward passes. PipeDream is an asynchronous approach, sending the output activations of each stage in a forward pass asynchronously to the next stage. Simultaneously, another mini-batch starts being processed during this communication. Similarly, the gradients, computed in each stage of a backward pass, are sent asynchronously to the previous stage, while another mini-batch starts processing. PipeDream leverages both pipeline and data parallelism together. Thus, the system allows multiple mini-batches as input for parallel training. This asynchronous communication of activations and gradients results in significant overlap of computation and communication as they operate on different mini-batches.


GPipe introduces a novel batch splitting mechanism where each mini-batch is splitted in micro-batches. The underlying pipeline algorithm is utilized to process the micro-batches of a mini-batch in parallel. In contrast to PipeDream, GPipe is synchronous implementing uni-directional training of DNNs, and it performs synchronous mini-batch gradient descent for DNN training, where the computed gradients are gathered across all micro-batches in a mini-batch and the updates are applied on the weights at the end of a mini-batch. 
The scheduling policy of GPipe limits its applicability merely to networks that can be expressed as a sequence of layers. Both GPipe and PipeDream exploit pipeline and data parallelism together. In contrast to PipeDream, periodic pipeline flushes are performed in GPipe after each mini-batch, in order to maintain single weight version at a time, making GPipe more memory efficient. There exists a single version of weight parameters for all the micro-batches in a mini-batch. 

There are two memory-efficient versions of PipeDream, namely PipeDream-2BW \cite{narayanan2021memory} and PipeDream-Flush \cite{narayanan2021memory}. PipeDream-2BW is based on a double buffered weight updates (2BW) technique, which reduces the memory footprint by reducing the number of active weight versions to two - one for already in-flight micro-batches (also known as shadow version) and the other for newly admitted micro-batches. In contrast to GPipe, it increases pipeline throughput by avoiding pipeline flushes. 
PipeDream-Flush achieves a smaller memory footprint than PipeDream-2BW by performing periodic pipeline flush at the cost of pipeline throughput. It maintains a single weight version at a time, which reduces the memory footprint. Both of them follows 1F1B scheduling, similar to PipeDream.

vPipe \cite{zhao2021v} is a system that introduces dynamic layer partitioning and memory management mechanisms for pipeline parallelism by searching for a plan of near-optimal layer partitioning and memory management. vPipe acts as a virtual layer between pipeline parallel systems such as PipeDream, GPipe, and so forth, and their execution engines such as PyTorch, Tensorflow, and among others. 
BPipe \cite{kim2023bpipe} is another approach in literature for achieving memory balance in pipeline parallelism. By using an activation balancing technique, BPipe allows all GPUs to use similar amounts of memory during training by transferring intermediate activations across them. 
Hippie \cite{10.1145/3472456.3472497} is a hybrid parallel training framework integrating pipeline parallelism and data parallelism to increase the throughput and scalability of large DNN training by hiding gradient communication. In addition, it provides a last-stage pipeline scheduling and recomputation mechanism for specific layers. 

PipePar \cite{zhang2023pipepar} is a model partitioning and task placement algorithm for pipeline parallel DNN training in heterogeneous GPU clusters. PipePar is based on dynamic programming with search space pruning that takes into consideration both the heterogeneity of GPUs and network bandwidth. 
 Koala \cite{tangkoala} is an automatic, end-to-end and  globally optimal searching technique for an efficient scheduling strategy with optimal flexibility and training efficiency. 
It facilitates solving the pipeline schedule development as a Binary-Tree-Traversing optimization problem.

Mario \cite{liu2025mario} is a pipeline optimizer technique that uniquely incorporates activation checkpointing into pipelines in order to achieve reduced and balanced memory footprint across devices. Additionally, it also consists of an automatic scheduler which can determine better parameter configurations. WeiPipe \cite{lin2025weipipe}, also known as weight-pipeline parallelism, is a weight-passing pipeline technique, which decreases communication costs across devices,
 and also ensures scalability 

DualPipe \cite{liu2024deepseek} is a bidirectional pipeline parallelism framework first introduced in DeepSeek V3 Technical Report \cite{liu2024deepseek}. DualPipe eliminates pipeline bubbles through dual-channel processing enabling simultaneous occurrence of forward-backward computation-communication phases and complete synchronization between forward and backward passes. It achieves optimized resource utilization, reduces idle time between processing stages and reduces memory footprint across all the devices. It uses an adaptive task scheduling method based on computational demands. 

MEPipe \cite{sun2025mepipe} introduces a slice-level scheduling method, named as Sequence Virtual Pipeline Parallelism (SVPP), for sequence pipeline parallelism. This method democratizes the training of LLMs to inexpensive accelerators with low-bandwidth interconnection by reducing memory footprint without increasing communication overhead across all devices. 
Zero Bubble \cite{qi2024zero} pipeline parallelism technique introduces a scheduling strategy that achieves almost zero pipeline bubbles under synchronous training mechanisms. In Zero Bubble, the backward computation is splitted into two parts - gradient computation for the input and that for the parameters. 

GraphPipe \cite{jeon2025graphpipe} is another pipeline parallelism technique that enables the system to identify the dependencies between different pipeline stages of a DNN by a directed acyclic graph. In contrast to traditional sequential pipeline parallelism (SPP), GraphPipe enables concurrent execution of computationally independent components of a DNN.

TiMePReSt (Time and Memory Efficient Pipeline Parallel DNN Training with Removed Staleness) \cite{dutta2024timeprest} is a memory-efficient pipeline parallelism based DNN training framework that addresses the above issue of stale weights. It has introduced an intra-batch scheduling technique, named as $n$F$1$B scheduling, that makes the framework more time-efficient.
\section{Methodology}
\label{sec:Methodology}
\begin{figure}[h]
  \centering
  \begin{subfigure}{0.45\textwidth}
    \centering
    \includegraphics[width=1.03\linewidth, height = 3cm]{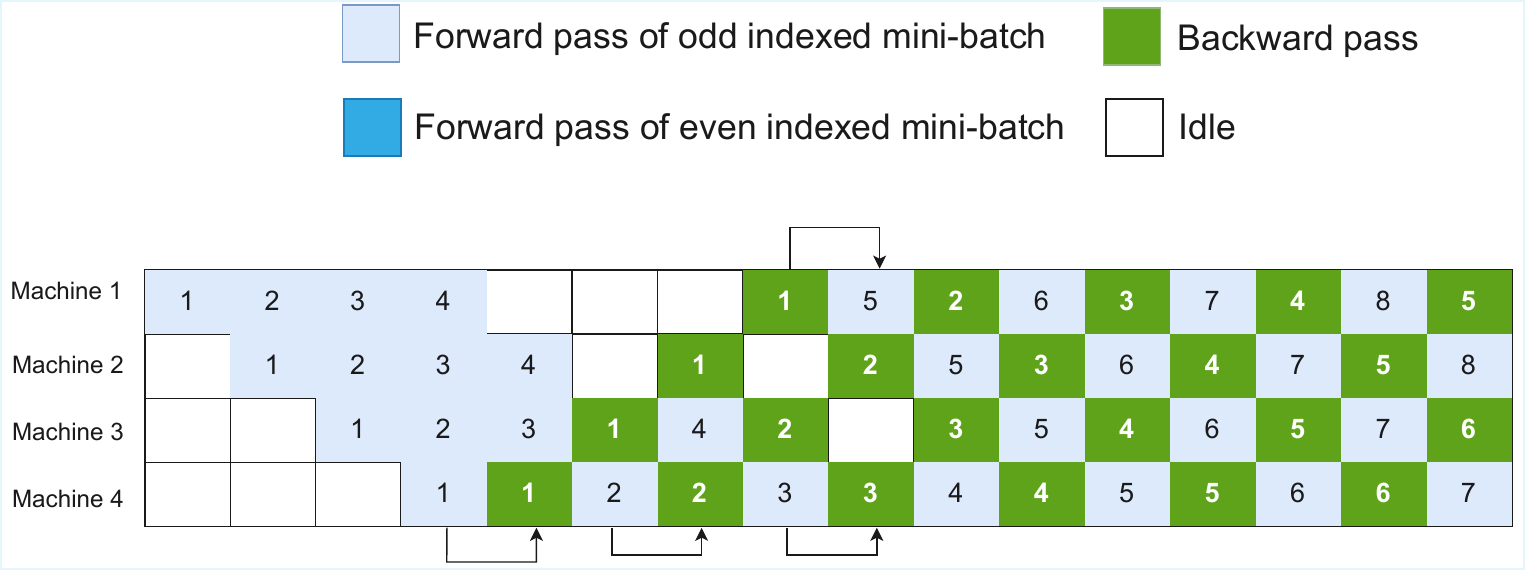}
    \caption{PipeDream}
    \label{fig:PipeDream}
  \end{subfigure}
  \begin{subfigure}{0.45\textwidth}
    \centering
    \includegraphics[width=1.05\linewidth]{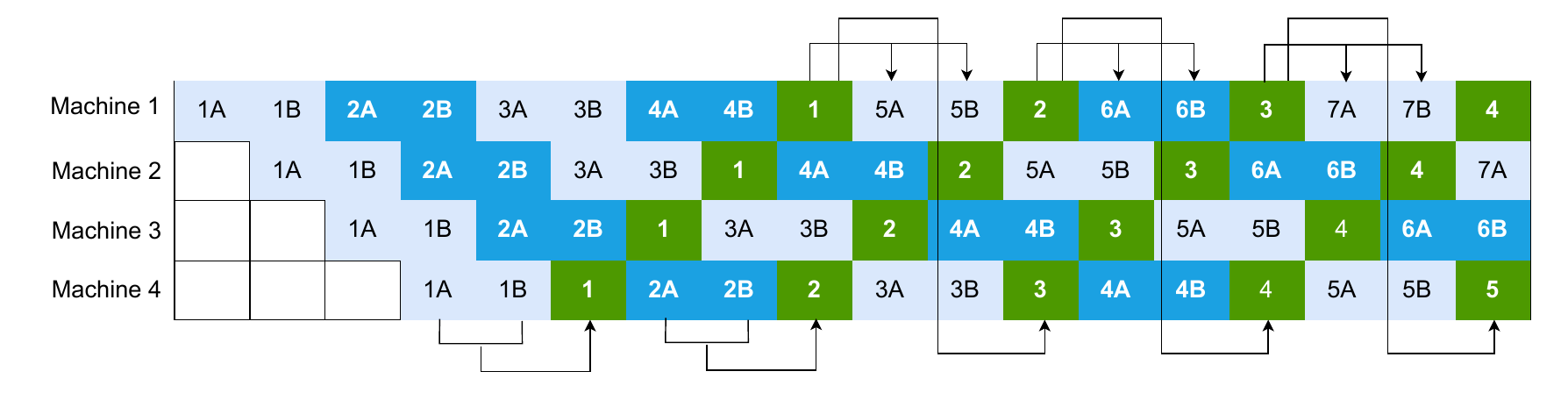}
    \caption{TiMePReSt}
    \label{fig:TiMePReSt_W4N2}
  \end{subfigure}
  \begin{subfigure}{0.45\textwidth}
  \centering
    \includegraphics[width=1.05\linewidth]{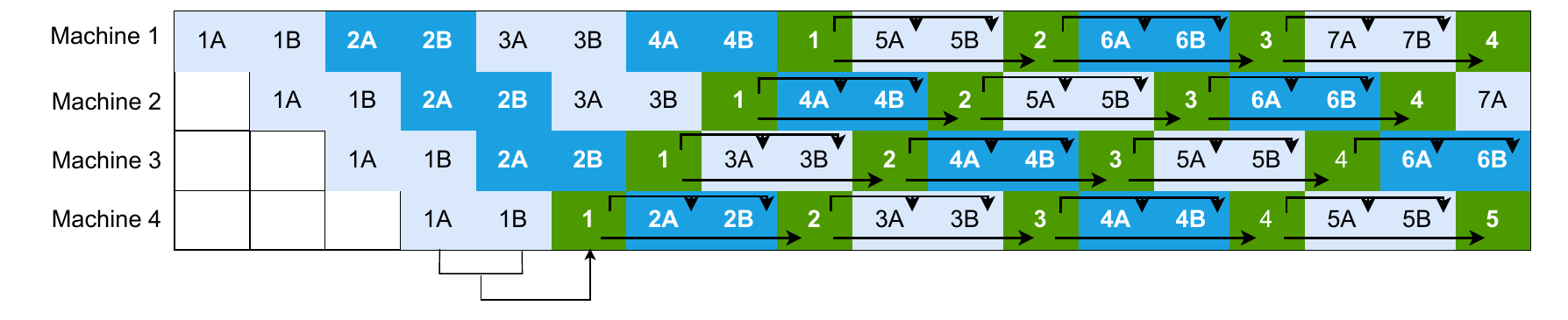}
    \caption{V-TiMePReSt}
    \label{fig:TiMePReSt-v}
  \end{subfigure}
  \begin{subfigure}{0.45\textwidth}
  \centering
    \includegraphics[width=1.03\linewidth]{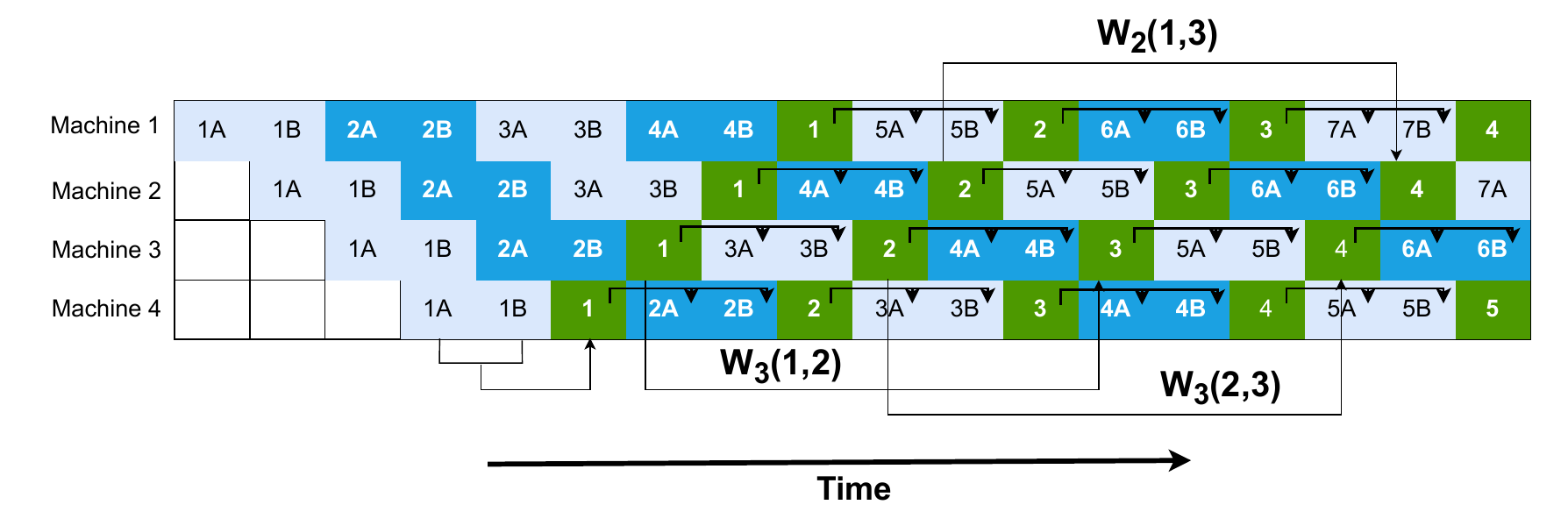}
    \caption{I-TiMePReSt}
    \label{fig:TiMePReSt-i}
  \end{subfigure}
  \caption{\textbf{PipeDream, TiMePReSt, V-TiMePReSt and I-TiMePReSt (top to bottom), the example pipeline parallel training schemes in an epoch with four workers each, are depicted above. TiMePReSt, V-TiMePReSt and I-TiMePReSt are having two micro-batches each. Numbers indicate mini-batch ID and alphabets indicate micro-batches. In contrast to Figures \ref{fig:TiMePReSt_W4N2}, \ref{fig:TiMePReSt-v} and \ref{fig:TiMePReSt-i}, PipeDream is having no distinction between odd and even indexed mini-batches.}}
  \label{fig:PipeDream, TiMePReSt, V-TiMePReSt and I-TiMePReSt}
\end{figure}

In this work, we address the problem of staleness of weights in stage-level, where each stage consists of one or more consecutive layers. In Figure \ref{fig:TiMePReSt_W4N2}, we observe that while the forward pass of mini-batch $4$ moves from stage $1$ to $2$, an update from mini-batch $1$ has already occurred. Thus, the version of weights on which the forward pass started becomes stale, but still is being used in the backward pass of the mini-batch 4. It indicates that the latest updated weights often cannot be utilized immediately in TiMePReSt \cite{dutta2024timeprest}. In V-TiMePReSt, we enable the system to utilize the updated weights as soon as they occur. Thus, it does not use the stale weights once the updated weights have been computed. The independence from stale weights eliminates the need for weight stashing, making the system more memory-efficient since the stale weights can be discarded immediately when updated version is available. 

However, the removal of weight stashing increases the number of epochs to converge since the training of a DNN does not depend solely on either stale or updated weights. Although the memory footprint is reduced with the elimination of weight stashing, however, in order to achieve a trade-off between the memory-efficiency and the number of epochs to converge, we introduce I-TiMePReSt. I-TiMePReSt is able to overcome the issue of staleness of weights and to achieve convergence within less number of epochs by making progress from the stale weights towards the actual updated weights. It neither uses the stale nor the actual updated weights, however, something in between them. Both V-TiMePReSt and I-TiMePReSt are comparable to TiMePReSt \cite{dutta2024timeprest} and PipeDream \cite{narayanan2019pipedream} depending on their performances in different DNNs and datasets we have used, although both have their own pros and cons.

Both V-TiMePReSt and I-TiMePReSt are pipeline parallelism based DNN training frameworks, leveraging pipelining and data parallelism together. As mentioned earlier, in pipeline parallelism based frameworks, the layers of a DNN are divided into sets of consecutive layers without overlapping, and these sets are distributed over the accelerators (GPUs) in a cluster. We develop the proposed frameworks on the same cluster of two machines, each having a single GPU, and distribute the DNNs under consideration in such a way that a balance is maintained between the memory consumptions in each node of the cluster. The nodes are indexed according to the assignment of the sets of layers. The index-wise first node gets the dataset on which the DNN is trained. Once the forward pass starts, each node communicates the output of the computations performed in it to the next node in the sequence. In the last node, the loss or prediction error is calculated, on which the backward pass starts and continued in the reverse direction of the forward pass. Similar to TiMePReSt \cite{dutta2024timeprest}, $n$ Forward $1$ Backward ($n$F$1$B) scheduling strategy is used in both V-TiMePReSt and I-TiMePReSt to schedule the forward and backward passes of multiple mini-batches, participating in parallel, in such a way, that no circular waiting for resources occurs between both the passes, and the computationally expensive nature of backward passes can be handled with an extra level of parallelism.       

Figures \ref{fig:TiMePReSt-v} and \ref{fig:TiMePReSt-i} represent examples of the proposed schemes of training of DNNs in V-TiMePReSt and I-TiMePReSt respectively. Similar to TiMePReSt, each mini-batch is split into some micro-batches in both the models. The forward passes of all the micro-batches perform in parallel, whereas only a single backward pass is performed collectively for all the micro-batches. More precisely, Figures \ref{fig:TiMePReSt-v} and \ref{fig:TiMePReSt-i} represent two example training schemes of V-TiMePReSt and I-TiMePReSt respectively, where each mini-batch is split into two micro-batches. The forward and backward passes are represented using blue and green colors, respectively, whereas light and dark blue colors are used for better visual differentiation of the forward passes of odd and even indexed mini-batches. Each box represents either forward pass of a micro-batch or backward pass of a mini-batch in a node at a time-point. The text inside each box consists of the mini-batch ID followed by the micro-batch ID (for forward pass only). The white boxes indicate idle states of the nodes.

\subsection{V-TiMePReSt}
\label{sec:V-TiMePReSt}

In this section, we discuss the design of V-TiMePReSt, a memory-efficient and zero-staleness variant of TiMePReSt.

\subsubsection*{\textbf{Model Architecture}}
\label{sec:Model Architecture of V-TiMePReSt}

V-TiMePReSt enables utilization of the latest updated weights in each stage of all the forward as well as backward passes. For example, in Figure \ref{fig:TiMePReSt-v}, we can see that the mini-batch $1$ uses the same version of weights in both forward and backward passes, since a single version of weights is available during the entire runtime of mini-batch $1$. However, in the last stage of the forward pass, mini-batch $2$ uses the weights updated by mini-batch $1$, whereas in the prior stages it does not. Similarly, mini-batch $3$ performs its forward pass on three different versions of weights - one in stages $1$ and $2$, another two in stages $3$ and $4$. Additionally, it searches for the latest updated weights and uses them while the backward pass starts, as in TiMePReSt. In V-TiMePReSt, each version of the weights on each machine is removed from the memory once the next version is available. We will now discuss in more detail how V-TiMePReSt handles problems such as staleness of weights and memory consumption during training in more detail.

\paragraph{\textbf{Challenge 1 - Staleness of Weights}}
The existing pipelining based DNN training systems in literature including TiMePReSt are not entirely free of stale weights. They suffer from varying degrees of staleness depending on the architecture of the model. As mentioned earlier, weight stashing is a frequently used approach to avoid weight staleness. To the best of our knowledge, TiMePReSt is the only model independent of horizontal weight stashing, and it incorporates the flexibility of utilizing different versions of weights in forward and backward passes on a mini-batch. However, during a pass it cannot utilize updated weights, if any. Thus, the learning does not become free of stale weights. 
\paragraph*{\textbf{Solution}}

In order to have a staleness-free pipeline parallelism framework for DNN training, V-TiMePReSt checks for the newer versions of weights even during the execution of a pass (forward or backward), rather than solely relying on the weights on which the computations of a pass have started. In other words, V-TiMePReSt checks for the updates in a frequent manner during forward pass (computation of the output of a job) as well as backward pass (analysis of the factors responsible for the result). V-TiMePReSt is not bound to use the updated weights only after a complete pass. Thus, it is flexible to measure the deviation of the computed output of a job with respect to the latest knowledge base, which was not available when the output was being computed. Thus, V-TiMePReSt achieves zero degree of staleness, which makes it unique among the existing models in literature. 
\paragraph{\textbf{Challenge 2 - GPU Memory Overhead}}
As discussed earlier, TiMePReSt and other existing models implement weight stashing to accommodate varying degrees of weight staleness, depending on the architecture of the pipelining framework. This approach allows different mini-batches to utilize different versions of the model weights, facilitating efficient pipeline execution. However, to support this mechanism, the stale knowledge base must be retained in memory until the associated forward or backward passes are fully completed, even though the updated knowledge base becomes available in the meantime.
The necessity of storing outdated versions of weights leads to significant memory overhead, as different values of the same weight parameters keep occupying memory for extended durations. As a consequence of the redundancy, excessive memory consumption makes it increasingly challenging to scale and train large-scale DNNs efficiently. The feasibility of training complex models having large number of layers is hindered due to the limited availability of GPU-memory, necessitating more memory-efficient strategies for training DNNs in distributed manner.  
\paragraph*{\textbf{Solution}} 

To address the memory inefficiency associated with weight stashing, we propose V-TiMePReSt, a novel pipeline parallelism based DNN training approach that eliminates the need to retain stale weight versions until their associated forward or backward passes are fully completed. Unlike conventional models that require maintaining outdated weight parameters throughout an epoch in pipeline execution, V-TiMePReSt enables DNNs to be trained with no utilization of stashed weights.

By eliminating redundant weight storage, V-TiMePReSt significantly optimizes memory utilization, reducing the overall GPU-memory footprint, whereas the existing pipeline parallelism strategies lack memory efficiency \cite{lin2025enhancing}. This enhancement not only improves the scalability and efficiency of the training process but also facilitates the training of large and complex DNN architectures that would otherwise be constrained by memory limitations. As a result, V-TiMePReSt provides a more resource-efficient solution for distributed DNN training, enabling high-performance computations without sacrificing model capacity or training effectiveness.

V-TiMePReSt blindly discards the stale weights once an updated version is available, regardless whether it has any significance in the training process of a DNN or not. This characteristics of V-TiMePReSt may be considered as a disadvantage.


\subsection{I-TiMePReSt}
\label{sec:I-TiMePReSt}

In this section, we discuss the design of I-TiMePReSt, a variant of TiMePReSt, that is able to maintain a trade-off among memory consumption, convergence speed and staleness of weights in an efficient way that considers the fact that all the versions of weights have a significance in the training process of a DNN. We have formulated a mathematical expression depicting the significance for I-TiMePReSt so that it can be computed numerically. 

\subsubsection*{\textbf{Model Architecture}}
\label{sec:Model Architecture of I-TiMePReSt}

As previously discussed, I-TiMePReSt is an advanced pipeline parallelism based DNN training framework. In TiMePReSt and V-TiMePReSt, it was observed that eliminating weight stashing mechanisms effectively reduces both weight staleness and memory consumption. However, it comes at the cost of the convergence speed in terms of the number of epochs required for convergence. In order to achieve a trade-off among weight staleness, memory efficiency, and convergence speed, we introduce I-TiMePReSt, a novel pipeline parallelism strategy for distributed DNN training. Unlike previous approaches, I-TiMePReSt does not completely rely on either stale weights or recently updated weights during training. Instead, it computes the significance of the stale weights based on the degree of staleness, and computationally progresses from stale weights toward updated weights, generating intermediate weight representations. These computed weights are then utilized during the backward propagation phase, ensuring an efficient training process while maintaining a controlled level of weight staleness.

We now derive an expression for intermediate weights in an epoch. Let $\textbf{W}_{i}(x|y)$, for a worker machine $i$, denote the updated weights computed using mini-batch $x$, which becomes stale once the updates using mini-batch $y$ is available. The term $\textbf{W}_{i}(x,y)$ stands for the intermediate weights between the stale weights $\textbf{W}_{i}(x|y)$ and the latest updated weights. Here $\textbf{W}_{i}(x,y)$ is computed based on $\textbf{W}_{i}(x|y)$ and the degree of staleness $\delta \geq 0$. The backward pass on a mini-batch is performed based on $\textbf{W}_{i}(x,y)$. We assume that the $\textbf{W}_{i}(x,y)$ is a combination of the stale weights $\textbf{W}_{i}(x|y)$ and other potential factors influencing the training process. 

For example, in Figure \ref{fig:TiMePReSt-i}, we can see that while mini-batch $4$ is participating in the backward pass on machine $2$, it does not avail the updates from mini-batch $3$ in order to maintain horizontal weight stashing. However, it also does not use the weights on which it performed forward pass, since it becomes stale over time. It calculates $\textbf{W}_{2}(1,3)$ based on the degree of staleness and the stale weights $\textbf{W}_{2}(1|3)$, where $\textbf{W}_{2}(1,3)$ is a combination of the stale weights $\textbf{W}_{2}(1|3)$ and other potential factors influencing the training process. The term $\textbf{W}_{2}(1,3)$ is assumed to be something intermediate between the stale weights $\textbf{W}_{2}(1|3)$ and the updated weights. 

Thus, an expression for $\textbf{W}_{i}(x,y)$ is derived as
\begin{equation}\label{eqn:weight update rule v0}
\begin{split}
    \textbf{W}_{i}(x,y) = (2 - \frac{1}{f(\delta)})\times\textbf{W}_{i}(x|y),
\end{split}
\end{equation}
where
\begin{equation}\label{eqn:significance of the stale weight component}
    f(\delta) \approx e^{-\lambda\delta} 
\end{equation}
The term $f(\delta) \in (0,1]$ is a function of $\delta$, and $\lambda > 0$ is a constant. The proof of the above is provided in Appendix \ref{sec:Technical Proof}.
\\
\paragraph{\textbf{Challenge 1 - Staleness of Weights}}

As discussed earlier, V-TiMePReSt completely eliminates the use of stale weights and entirely removes their influence on the training process of a DNN. While this approach effectively mitigates weight staleness and optimizes memory efficiency, it comes at the cost of the number of epochs required for convergence. The absence of stale weight contributions may slow down the overall training speed, requiring additional iterations to achieve good model performance. 
\paragraph*{\textbf{Solution}}

As previously discussed, I-TiMePReSt is designed based on the assumption that stale weights play a significant role in the training process of a DNN, although they are not the only contributing factor. In this paper, the comparisons of PipeDream, TiMePReSt, and V-TiMePReSt have demonstrated that eliminating weight staleness leads to an increase in the number of epochs required for convergence.

The primary objective of I-TiMePReSt is to maintain a trade-off between memory consumption and convergence speed. Unlike existing models, which are solely dependent on either stale weights or updated weights, I-TiMePReSt adopts a more comprehensive approach by considering both as contributory factors. Additionally, it incorporates a flexible framework that allows inclusion of other potentially unknown factors influencing the training process. This unique approach differentiates I-TiMePReSt from the previous methods, and enhances its adaptability in distributed DNN training environments.
\paragraph{\textbf{Challenge 2 - GPU Memory Overhead}}

Among the existing models, including PipeDream, TiMePReSt, and V-TiMePReSt, V-TiMePReSt framework demonstrates the most efficient GPU-memory utilization. It is primarily due to its complete elimination of weight stashing, which significantly reduces memory overhead during training. However, this efficiency is achieved at the cost of entirely disregarding the contribution of stale weights in the training process of a DNN.

By fully removing the influence of stale weights, V-TiMePReSt minimizes memory consumption but affects the training dynamics, potentially impacting model convergence speed in terms of number of epochs. While this approach ensures better resource allocation, it may necessitate a larger number of training epochs to achieve comparable performance. Thus, while V-TiMePReSt excels in memory efficiency, its design lacks the trade-off between optimal usage of computational resources and convergence behavior, which must be carefully considered in large-scale distributed DNN training. In contrast, TiMePReSt keeps weight staleness to a significant extent. Thus, TiMePReSt is superior than V-TiMePReSt in terms of convergence speed, however, not in terms of memory consumption. 
\paragraph*{\textbf{Solution}}

To effectively balance the trade-off between GPU-memory consumption and model convergence speed, we propose I-TiMePReSt, a novel pipeline parallelism based DNN training framework. Unlike existing approaches that either prioritize memory efficiency by eliminating weight stashing or retain stale weights at the cost of increased memory usage, I-TiMePReSt adopts a more adaptive strategy to optimize both aspects simultaneously.

By leveraging an intermediate weight computation mechanism, I-TiMePReSt ensures that the training process benefits from the contribution of stale weights while preventing excessive memory overhead. It enables the framework to achieve efficient resource utilization without significantly increasing the number of epochs required for convergence. As a result, I-TiMePReSt provides a memory-efficient solution for distributed DNN training, addressing key limitations of existing models while maintaining training efficiency.
\paragraph{\textbf{Challenge 3 - High Training Time}}

V-TiMePReSt exhibits the slowest convergence speed among all the models mentioned earlier. It is primarily due to its complete elimination of weight staleness and its influence on the training process of a DNN. While removing stale weights reduces memory overhead, it also disrupts the natural learning speed by discarding potentially useful information embedded in stale updates. As a result, V-TiMePReSt requires a larger number of training epochs to achieve convergence compared to other models that incorporate stale weights to some extent. This pattern highlights the impact of stale weight utilization on training time. More precisely, it suggests that a complete removal, while beneficial for memory optimization, may negatively affect convergence speed. 
\paragraph*{\textbf{Solution}}

To mitigate slow convergence observed in V-TiMePReSt while maintaining memory efficiency, we have introduced I-TiMePReSt to achieve a trade-off between convergence speed and GPU-memory consumption. Unlike V-TiMePReSt, which fully eliminates stale weights and consequently slows down convergence, I-TiMePReSt incorporates a more adaptive approach that leverages both stale and updated weights in a computationally efficient manner. The proposed strategy of handling weight updates allows I-TiMePReSt to not only enhance convergence speed compared to V-TiMePReSt, but also to outperform TiMePReSt in terms of faster model convergence while maintaining comparable memory efficiency. As a result, I-TiMePReSt provides an efficient platform for distributed DNN training, addressing key limitations of the existing models.

\section{Evaluation}
\label{sec:Evaluation}

\paragraph{Time needed to achieve target accuracy}

As discussed earlier, V-TiMePReSt allows different weights in successive stages of a DNN, which makes the system capable of using the latest updated weights in each stage, resulting in the framework completely free of staleness. V-TiMePReSt achieves the flexibility at the expense of weight stashing. Thus, it requires more clock time to achieve a target accuracy compared to TiMePReSt and I-TiMePReSt. In contrast, I-TiMePReSt requires less time compared to V-TiMePReSt, TiMePReSt, and PipeDream due to its combined staleness reduction and weight stashing strategy. We compare V-TiMePReSt, I-TiMePReSt, TiMePReSt, and PipeDream with respect to time-vs-accuracy for VGG-16 and ResNet-50 training on CIFAR-100 and Tiny-ImageNet-200 image classification datasets using a cluster consisting of two machines having a single GPU each. One is NVIDIA Quadro RTX 6000 with 24 GB of GPU-memory, another is NVIDIA GeForce RTX 2080 with 12 GB of GPU-memory. 

\begin{figure}[h]
  \begin{subfigure}{.23\textwidth}
    \includegraphics[width=\linewidth]{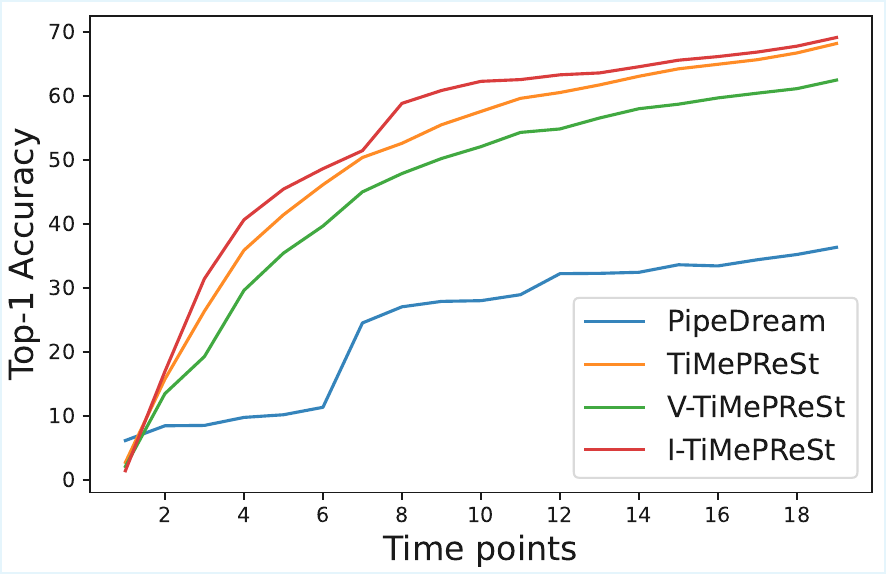}
    \caption{Top-1 accuracy vs time}
    \label{fig:Top-1 accuracy to time_cifar100_VGG-16}
  \end{subfigure}
  \begin{subfigure}{.23\textwidth}
    \includegraphics[width=\linewidth]{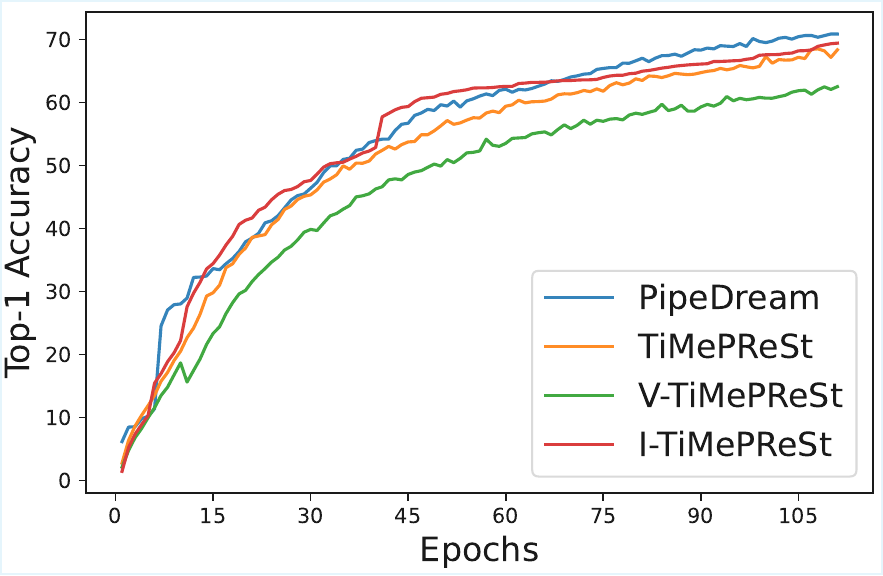}
    \caption{Top-1 accuracy vs epoch}
    \label{fig:Top-1 accuracy to epoch_cifar100_VGG-16}
  \end{subfigure}
  \begin{subfigure}{.23\textwidth}
    \includegraphics[width=\linewidth]{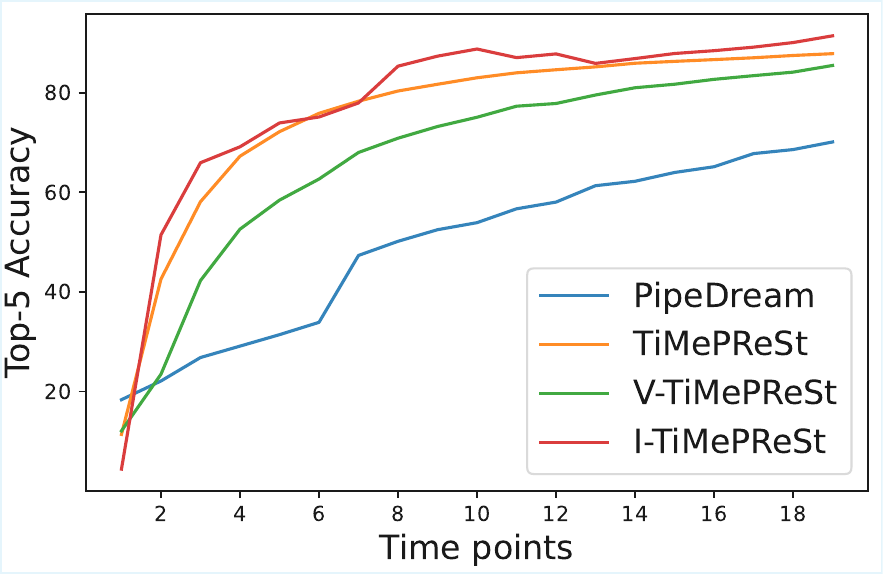}
    \caption{Top-5 accuracy vs time}
    \label{fig:Top-5 accuracy to time_cifar100_VGG-16}
  \end{subfigure}
  \begin{subfigure}{.23\textwidth}
    \includegraphics[width=\linewidth]{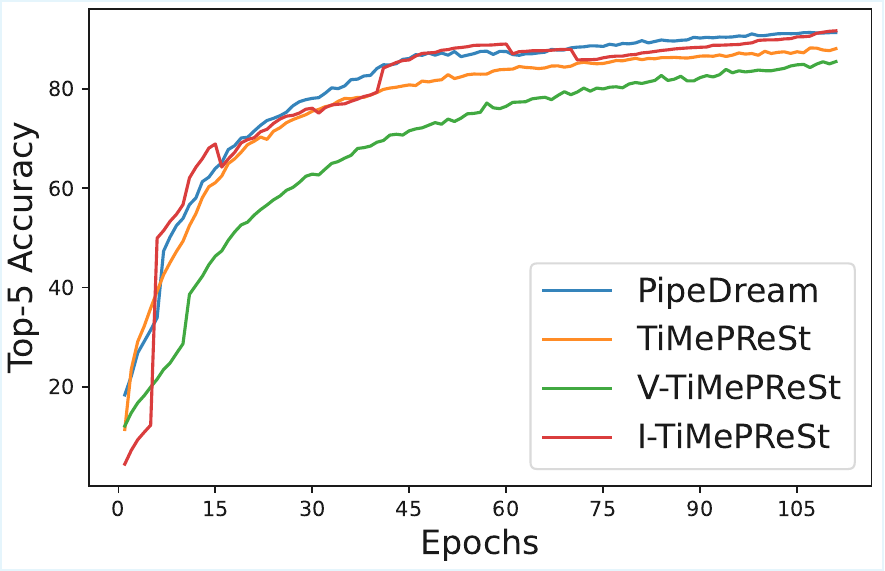}
    \caption{Top-5 accuracy vs epoch}
    \label{fig:Top-5 accuracy to epoch_cifar100_VGG-16}
  \end{subfigure}
  \begin{subfigure}{.23\textwidth}
    \includegraphics[width=\linewidth]{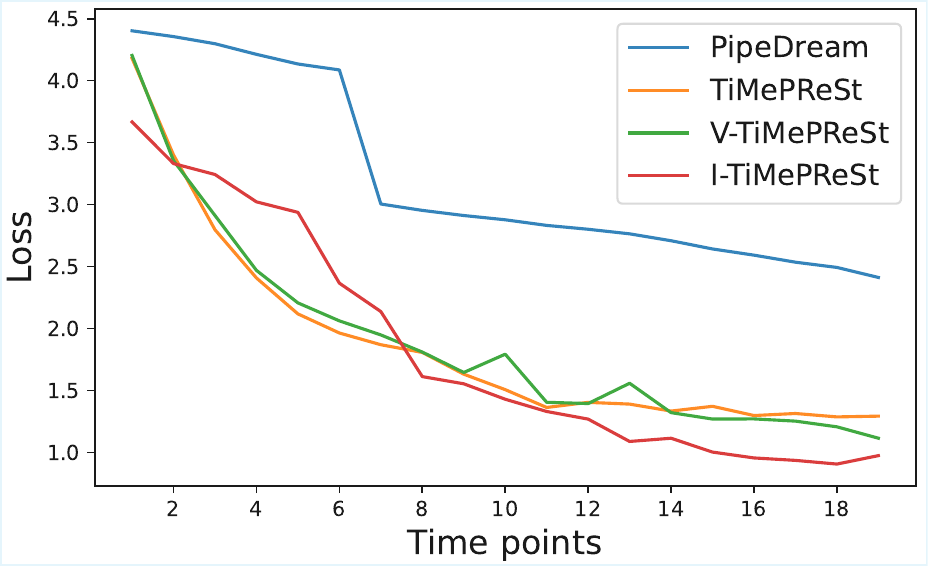}
    \caption{Loss vs time}
    \label{fig:Loss comparison_time_cifar100_VGG-16}
  \end{subfigure}
  \hspace{0.6em}
   \begin{subfigure}{.23\textwidth}
    \includegraphics[width=\linewidth]{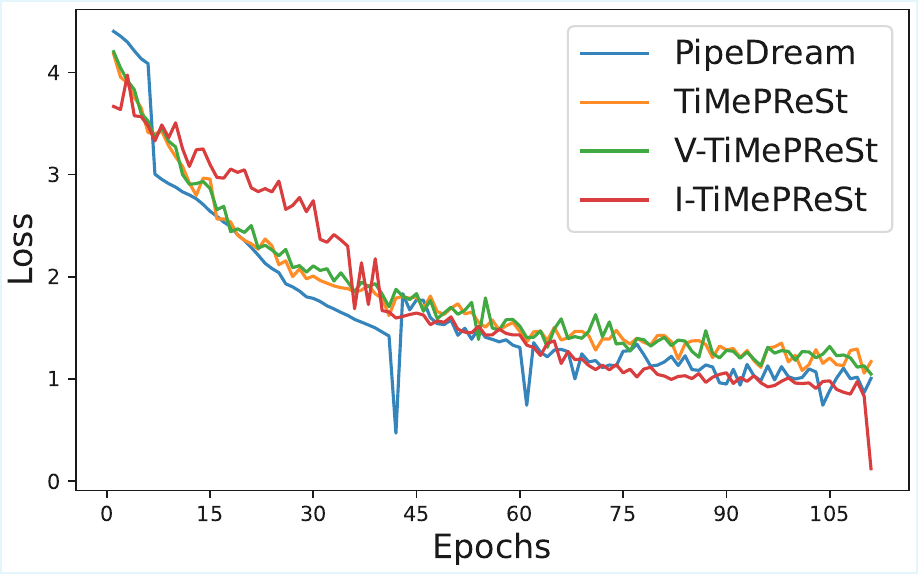}
    \caption{Loss vs epoch}
    \label{fig:Loss comparison_epoch_cifar100_VGG-16}
  \end{subfigure}
  \caption{\textbf{Performance comparison of V-TiMePReSt, I-TiMePReSt, TiMePReSt, and PipeDream platforms for VGG-16 training on CIFAR-100 dataset based on accuracy and loss achieved over time and epochs. Figures \ref{fig:Top-1 accuracy to time_cifar100_VGG-16}, \ref{fig:Top-5 accuracy to time_cifar100_VGG-16} and \ref{fig:Loss comparison_time_cifar100_VGG-16} represent the plots with respect to time, whereas Figures \ref{fig:Top-1 accuracy to epoch_cifar100_VGG-16}, \ref{fig:Top-5 accuracy to epoch_cifar100_VGG-16} and \ref{fig:Loss comparison_epoch_cifar100_VGG-16} represent the plots with respect to epochs (statistical efficiency).}}
  \label{fig:Performance Comparison of V-TiMePReSt, I-TiMePReSt, TiMePReSt and PipeDream (VGG-16 on CIFAR-100)}
\end{figure}

\begin{figure}[h]
  \begin{subfigure}{.23\textwidth}
  \centering
    \includegraphics[width=\linewidth]{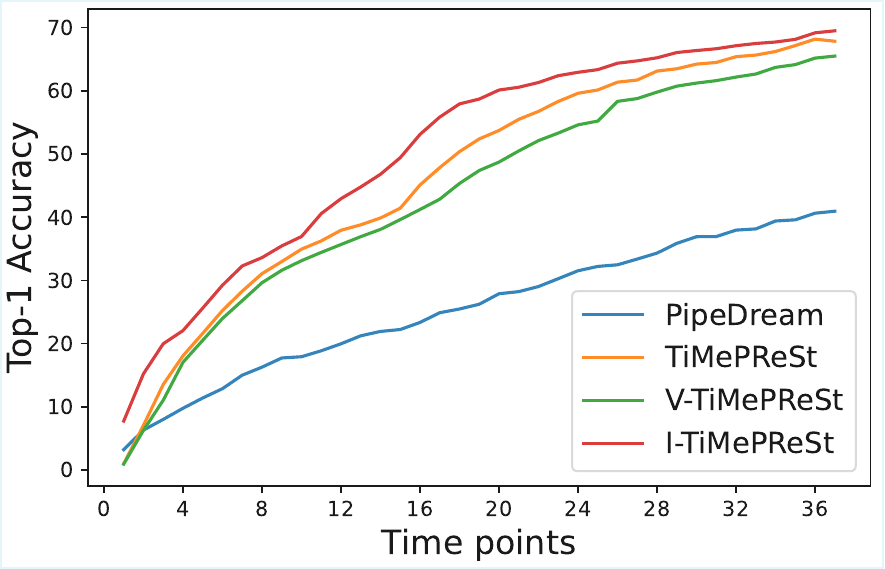}
    \caption{Top-1 accuracy vs time}
    \label{fig:Top-1 accuracy to time_tiny_ImageNet_VGG-16}
  \end{subfigure}
  \begin{subfigure}{.23\textwidth}
  \centering
    \includegraphics[width=\linewidth]{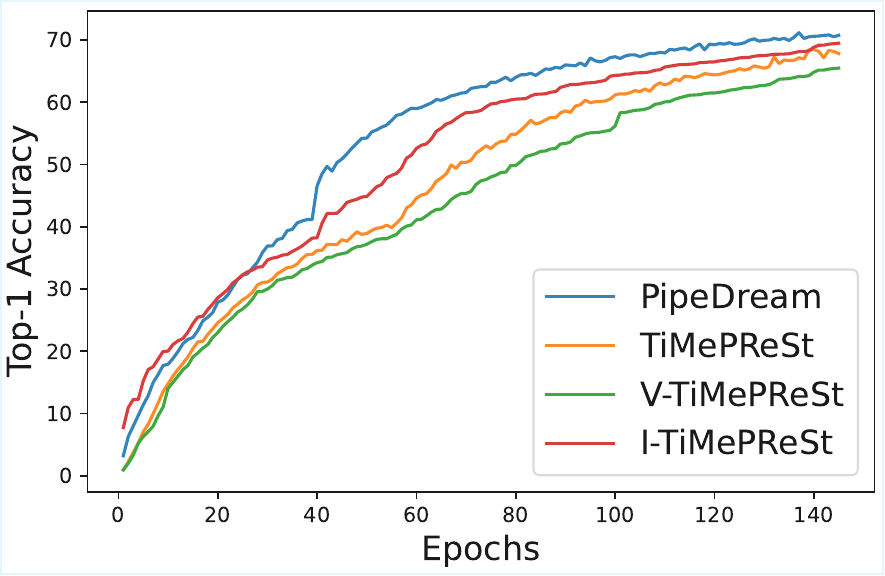}
    \caption{Top-1 accuracy vs epoch}
    \label{fig:Top-1 accuracy to epoch_TinyImageNet200_VGG-16}
  \end{subfigure}
  \begin{subfigure}{.23\textwidth}
  \centering
    \includegraphics[width=\linewidth]{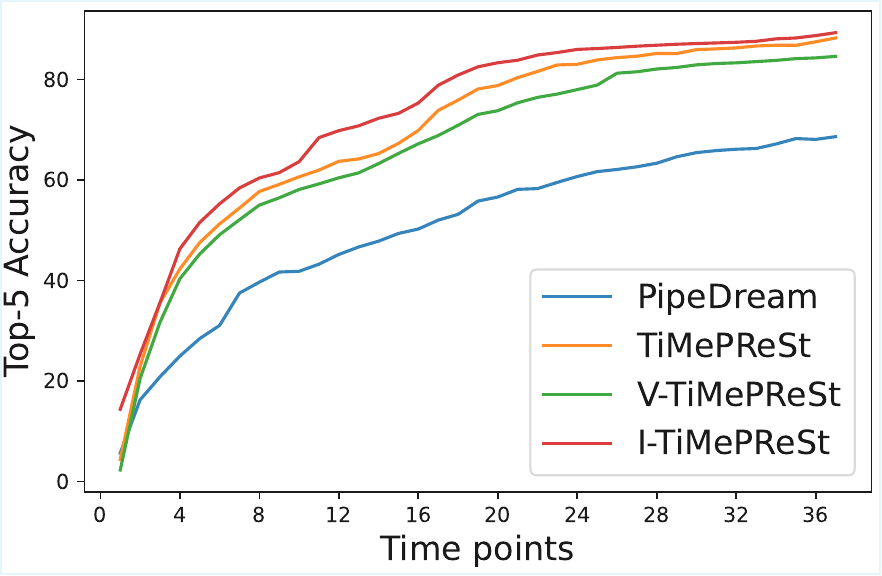}
    \caption{Top-5 accuracy vs time}
    \label{fig:Top-5 accuracy to time_tiny_ImageNet_VGG-16}
  \end{subfigure}
  \begin{subfigure}{.23\textwidth}
  \centering
    \includegraphics[width=\linewidth]{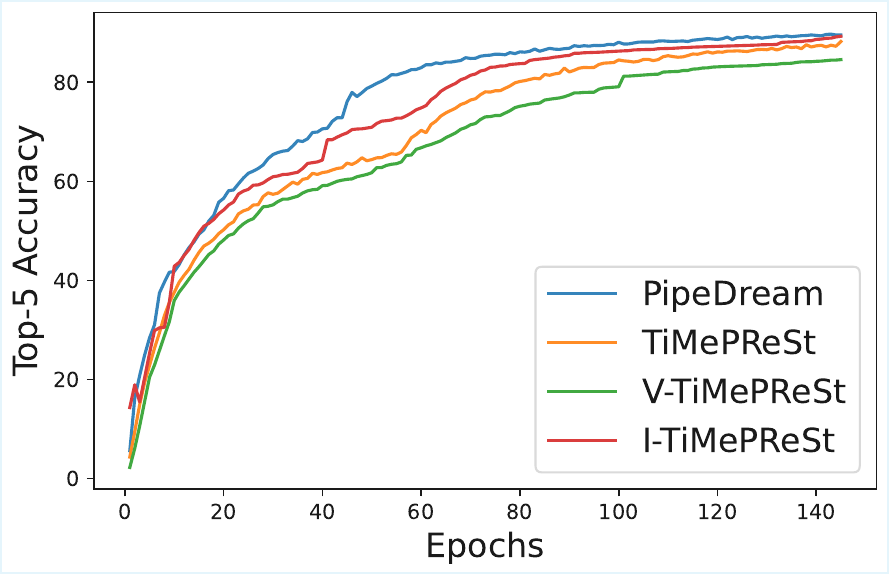}
    \caption{Top-5 accuracy vs epoch}
    \label{fig:Top-5 accuracy to epoch_TinyImageNet200_VGG-16}
  \end{subfigure}
  \begin{subfigure}{.23\textwidth}
  \centering
    \includegraphics[width=\linewidth]{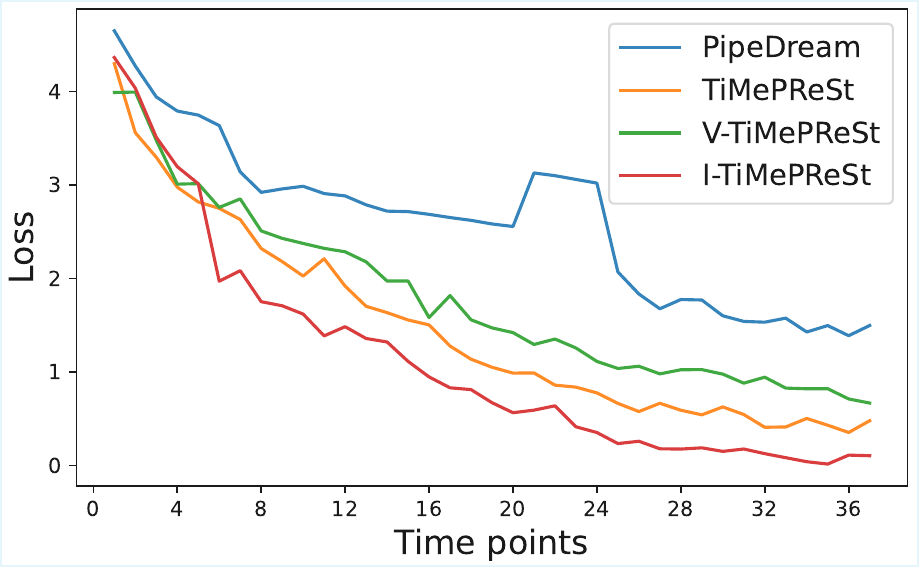}
    \caption{Loss vs time}
    \label{fig:Loss comparison_time_tiny_ImageNet_VGG-16}
  \end{subfigure}
  \hspace{0.6em}
  \begin{subfigure}{.23\textwidth}
  \centering
    \includegraphics[width=\linewidth]{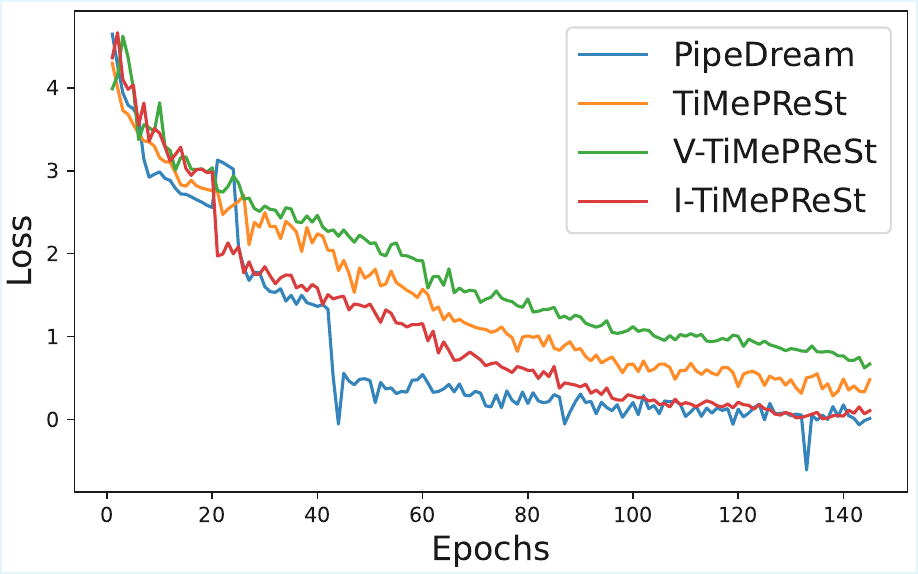}
    \caption{Loss vs epoch}
    \label{fig:Loss comparison_epoch_TinyImageNet200_VGG-16}
  \end{subfigure}
  \caption{\textbf{Performance comparison of V-TiMePReSt, I-TiMePReSt, TiMePReSt, and PipeDream platforms for VGG-16 training on Tiny-ImageNet-200 dataset based on accuracy and loss achieved over time and epochs. Figures \ref{fig:Top-1 accuracy to time_tiny_ImageNet_VGG-16}, \ref{fig:Top-5 accuracy to time_tiny_ImageNet_VGG-16} and \ref{fig:Loss comparison_time_tiny_ImageNet_VGG-16} represent the plots with respect to time, whereas Figures \ref{fig:Top-1 accuracy to epoch_TinyImageNet200_VGG-16}, \ref{fig:Top-5 accuracy to epoch_TinyImageNet200_VGG-16} and \ref{fig:Loss comparison_epoch_TinyImageNet200_VGG-16} represent the plots with respect to epochs (statistical efficiency).}}
  \label{fig:Performance Comparison of V-TiMePReSt, I-TiMePReSt, TiMePReSt and PipeDream (VGG-16 on Tiny_ImageNet)}
\end{figure}

Figures \ref{fig:Top-1 accuracy to time_cifar100_VGG-16} and \ref{fig:Top-5 accuracy to time_cifar100_VGG-16} show that V-TiMePReSt reaches the target top-1 and top-5 accuracy slower than I-TiMePReSt, TiMePReSt, but faster than PipeDream, in case of VGG-16 on CIFAR-100. Likewise, Figures \ref{fig:Top-1 accuracy to time_tiny_ImageNet_VGG-16} and \ref{fig:Top-5 accuracy to time_tiny_ImageNet_VGG-16} show similar results for VGG-16 on Tiny-ImageNet-200. Similar performance is also achieved for ResNet-50 as shown in Figures \ref{fig:Top-1 accuracy to time_cifar100_ResNet50} and \ref{fig:Top-5 accuracy to time_cifar100_ResNet50} in Appendix \ref{sec:Supplementary Experimental Results}. VGG-16 is a type of CNN (Convolutional Neural Network) consisting of sixteen weight layers or learnable parameter layers, whereas ResNet-50 is a CNN architecture having fifty layers. The capability of V-TiMePReSt to achieve target accuracy faster than PipeDream for both of them proves its scalability over PipeDream for larger DNNs. However, I-TiMePReSt achieves the highest scalability compared to the others.
\begin{figure}[h]
  \centering
  \begin{subfigure}{.25\textwidth} 
    \centering
    \includegraphics[width=\linewidth, height=7cm]{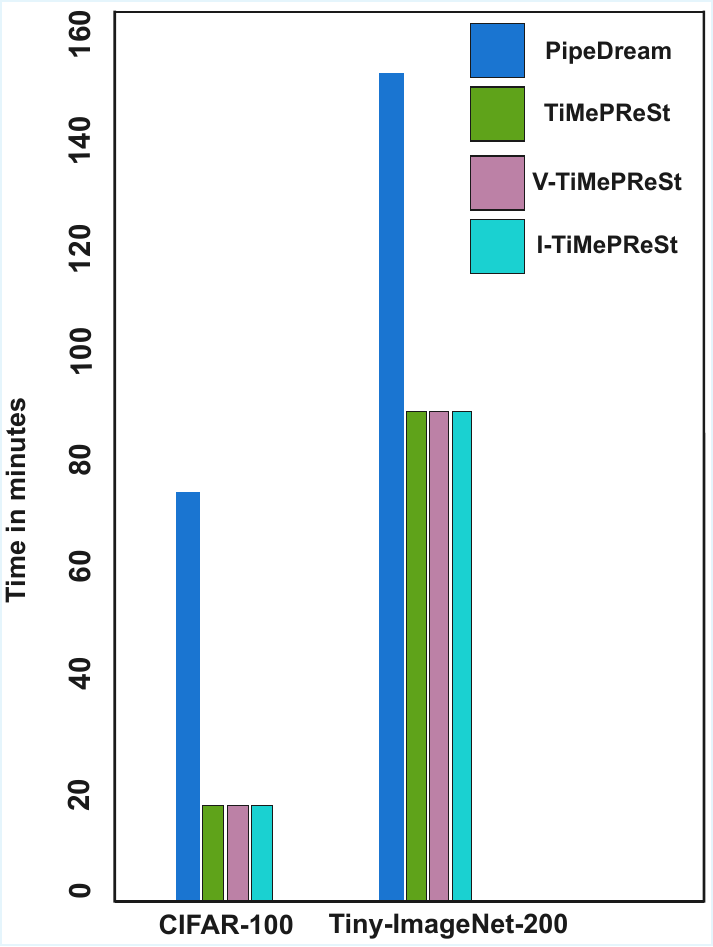}
    \caption{Hardware Efficiency (Time per epoch)}
    \label{fig:Time per epoch VGG-16}
  \end{subfigure}%
  \hfill
  \begin{subfigure}{.22\textwidth}
    \centering
    \includegraphics[width=\linewidth, height=6cm]{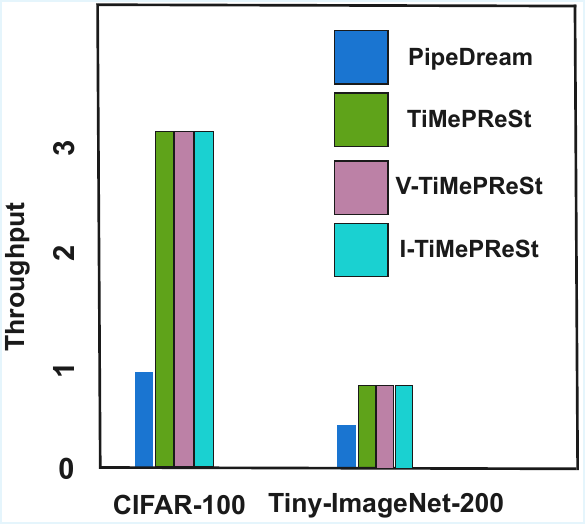}
    \caption{Throughput (epochs/hr)}
    \label{fig:Throughput VGG-16}
  \end{subfigure}
  \caption{\textbf{Comparison among V-TiMePReSt, I-TiMePReSt, TiMePReSt and PipeDream platforms based on hardware efficiency and throughput for training VGG-16 on CIFAR-100 and Tiny-ImageNet-200 datasets.}}
  \label{fig:Comparison between V-TiMePReSt, I-TiMePReSt, TiMePReSt and PipeDream (VGG-16 on Tiny-ImageNet-200), based on hardware efficiency and throughput}
\end{figure}
\begin{figure}[h]
  \centering
   \includegraphics[width=1.0\linewidth]{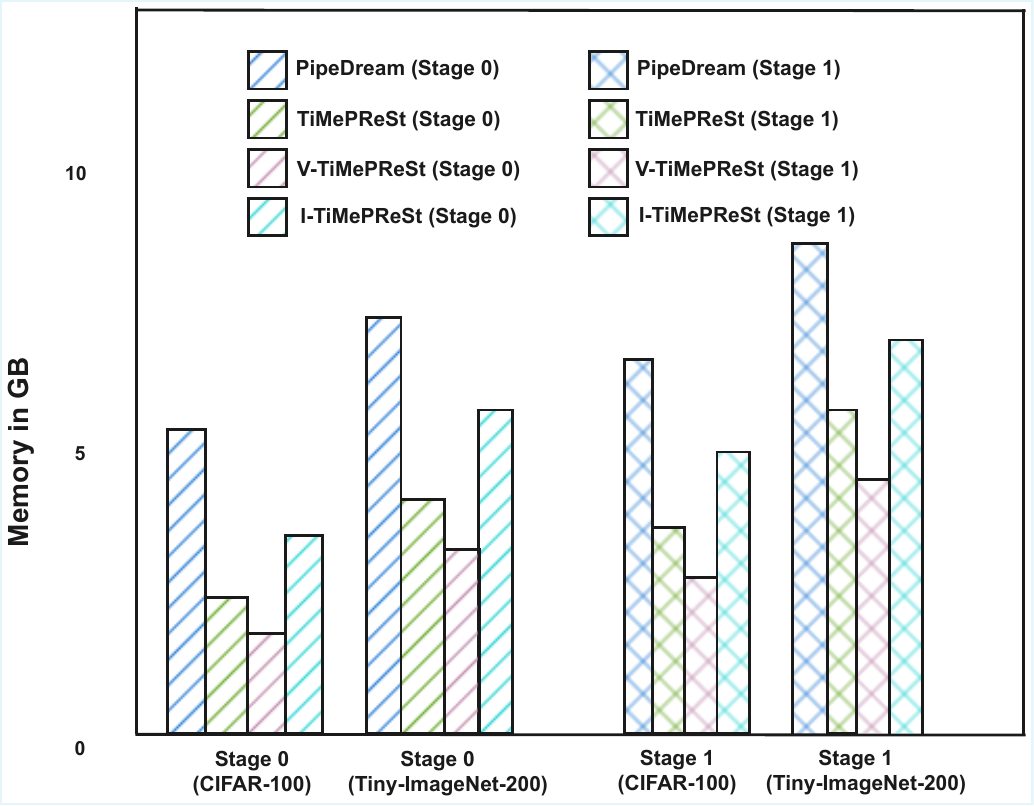}
  \caption{\textbf{Memory footprint per stage using two worker nodes for training VGG-16 on CIFAR-100 and Tiny-ImageNet-200 image classification dataset, for V-TiMePReSt, I-TiMePReSt, TiMePReSt and PipeDream platforms.}}
  \label{fig:Memory consumption comparison for VGG-16}
\end{figure}
\paragraph{Accuracy reached after equal execution time}

Figures \ref{fig:Top-1 accuracy to time_cifar100_VGG-16}, \ref{fig:Top-5 accuracy to time_cifar100_VGG-16}, \ref{fig:Top-1 accuracy to time_tiny_ImageNet_VGG-16}, and \ref{fig:Top-5 accuracy to time_tiny_ImageNet_VGG-16} show that V-TiMePReSt achieves less prediction accuracy after equal execution time compared to I-TiMePReSt and TiMePReSt, but more than PipeDream for VGG-16 on CIFAR-100 and Tiny-ImageNet-200 datasets. On the other hand, I-TiMePReSt achieves better accuracy compared to all. Similar performance is also achieved for ResNet-50 as shown in Figures \ref{fig:Top-1 accuracy to time_cifar100_ResNet50} and \ref{fig:Top-5 accuracy to time_cifar100_ResNet50} in Appendix \ref{sec:Supplementary Experimental Results}. The capability of V-TiMePReSt to achieve a higher accuracy than PipeDream after equal execution time for both VGG-16 and ResNet-50 proves its scalability once again over PipeDream for larger DNNs. However, I-TiMePReSt outperforms the others in terms of scalability.

\paragraph{Reduction of loss after equal execution time}

Figures \ref{fig:Loss comparison_time_cifar100_VGG-16} and \ref{fig:Loss comparison_time_tiny_ImageNet_VGG-16} show that V-TiMePReSt performs worse than I-TiMePReSt and TiMePReSt, but better than PipeDream in terms of reduction of loss after equal execution time. In other words, I-TiMePReSt is able to reduce the loss faster than the others for VGG-16 on CIFAR-100 and Tiny-ImageNet-200 datasets. Figure \ref{fig:Loss comparison_time_cifar100_ResNet50} in Appendix \ref{sec:Supplementary Experimental Results} shows the performances of the models for ResNet-50 on CIFAR-100. All these Figures \ref{fig:Loss comparison_time_cifar100_VGG-16},\ref{fig:Loss comparison_time_tiny_ImageNet_VGG-16} and \ref{fig:Loss comparison_time_cifar100_ResNet50} support the fact that I-TiMePReSt is the most scalable among the others, whereas V-TiMePReSt is more scalable compared to PipeDream. 





\paragraph{Number of epochs per hour}
Figures \ref{fig:Time per epoch VGG-16} and \ref{fig:Throughput VGG-16} report the fact that both V-TiMePReSt and I-TiMePReSt are able to perform similar number of epochs to TiMePReSt in an hour, which is much higher than PipeDream, since both V-TiMePReSt and I-TiMePReSt employs the micro-batch based training following $n$F$1$B scheduling similar to TiMePReSt. Figures \ref{fig:Time per epoch ResNet-50} and \ref{fig:Throughput ResNet-50} in Appendix \ref{sec:Supplementary Experimental Results} support the same. As we know, the number of epochs to be performed per hour is known as \textbf{throughput} \cite{narayanan2019pipedream}. Thus, we can say that both V-TiMePReSt and I-TiMePReSt are better than PipeDream and similar to TiMePReSt in terms of throughput for VGG-16 on both CIFAR-100 and Tiny-ImageNet-200 datasets. Instead of using epochs or actual clock time, all the plots in Figures \ref{fig:Performance Comparison of V-TiMePReSt, I-TiMePReSt, TiMePReSt and PipeDream (VGG-16 on CIFAR-100)} and \ref{fig:Performance Comparison of V-TiMePReSt, I-TiMePReSt, TiMePReSt and PipeDream (VGG-16 on Tiny_ImageNet)} use time points on the x-axis for improved comparability. The time interval between two successive time points is nothing but the single-epoch time for any of the models - PipeDream, TiMePReSt, V-TiMePReSt, or I-TiMePReSt, whichever is larger. Since epoch time varies with datasets and DNNs under consideration, the duration of a single time-interval also varies with datasets.

\paragraph{Hardware efficiency, statistical efficiency and training time}
Figure \ref{fig:Comparison between V-TiMePReSt, I-TiMePReSt, TiMePReSt and PipeDream (VGG-16 on Tiny-ImageNet-200), based on hardware efficiency and throughput} demonstrates that both V-TiMePReSt and I-TiMePReSt are comparable to TiMePReSt, and more efficient than PipeDream in terms of time per epoch, which is also known as \textit{hardware efficiency} \cite{narayanan2019pipedream}, for VGG-16 on both the datasets mentioned earlier. However, in Figures \ref{fig:Top-1 accuracy to epoch_cifar100_VGG-16}, \ref{fig:Top-5 accuracy to epoch_cifar100_VGG-16} and \ref{fig:Loss comparison_epoch_cifar100_VGG-16}, we can see that I-TiMePReSt outperforms V-TiMePReSt and TiMePReSt in terms of \textit{statistical efficiency} (number of epochs needed to achieve a particular accuracy) \cite{narayanan2019pipedream} for VGG-16 with CIFAR-100, while giving comparable performance to PipeDream. Similar kind of results are demonstrated in Figures \ref{fig:Top-1 accuracy to epoch_TinyImageNet200_VGG-16}, \ref{fig:Top-5 accuracy to epoch_TinyImageNet200_VGG-16} and \ref{fig:Loss comparison_epoch_TinyImageNet200_VGG-16} (VGG-16 with Tiny-ImageNet-200), Figures \ref{fig:Top-1 accuracy to epoch_cifar100_ResNet50}, \ref{fig:Top-5 accuracy to epoch_cifar100_ResNet50} and \ref{fig:Loss comparison_epoch_cifar100_ResNet50} (ResNet-50 with CIFAR-100) in Appendix \ref{sec:Supplementary Experimental Results}, and Figures \ref{fig:Top-1 accuracy to epoch_TinyImageNet200_ResNet50}, \ref{fig:Top-5 accuracy to epoch_TinyImageNet200_ResNet50} and \ref{fig:Loss comparison_epoch_TinyImageNet200_ResNet50} (ResNet-50 with Tiny-ImageNet-200) in Appendix \ref{sec:Supplementary Experimental Results}. Thus, the total training time of a DNN is also shorter in I-TiMePReSt framework than in V-TiMePReSt and TiMePReSt.


\paragraph{Memory Overhead}
Since V-TiMePReSt completely removes weight stashing in order to become fully staleness-aware, it suffers from more statistical efficiency problem. However, at the expense of statistical efficiency, V-TiMePReSt achieves the minimum memory consumption among all other models. Although I-TiMePReSt is also staleness-aware, the memory consumption is compromised. The reason is that it keeps the stale weights till the improved weights (towards the actual updated weights) are computed for the last mini-batch which used the particular stale weights in forward pass. I-TiMePReSt consumes more memory compared to TiMePReSt, since TiMePReSt is free of horizontal weight stashing. On the other hand, I-TiMePReSt is not completely free of stashing instead of being staleness-aware. Figure \ref{fig:Memory consumption comparison for VGG-16} demonstrates the GPU memory requirements of all the models for VGG-16 on both the datasets.
\section{Conclusion}
\label{sec:Conclusion}

In this work, we have explored the challenges associated with pipeline parallelism-based DNN training, particularly in the context of weight staleness, memory efficiency, and convergence speed. We have introduced two such frameworks having architectural similarity, such as V-TiMePReSt and I-TiMePReSt. Each of them offers different trade-offs in addressing these challenges. While V-TiMePReSt achieves the lowest memory consumption by completely eliminating stale weights, it does so at the cost of significantly reducing convergence speed. On the other hand, I-TiMePReSt effectively balances memory consumption and convergence speed. By incorporating an intermediate weight computation mechanism, I-TiMePReSt ensures that the training process is benefited from stale weights without excessive memory overhead, achieving a trade-off between training efficiency and computational resource utilization. Experimental results demonstrate that I-TiMePReSt outperforms V-TiMePReSt as well as TiMePReSt in terms of convergence speed while also surpassing PipeDream sometimes. In terms of memory efficiency, V-TiMePReSt outperforms all these models, while I-TiMePReSt is comparable to TiMePReSt. Moreover, both I-TiMePReSt is the most scalable among all the other models mentioned earlier, whereas V-TiMePReSt is also more scalable than PipeDream. Thus, each of them can be considered as an effective solution for distributed DNN training in different scenario. In addition, we have formulated an expression depicting the significance of each stale weight based on the degree of staleness. To the best of our knowledge, the literature does not contain any such quantitative description of the significance of the stale weight. Morever, in case of I-TiMePReSt, a computational approach has been developed for intermediate weight calculation between stale and updated weights, which is also unique in literature.  


\bibliographystyle{ieeetr}
\bibliography{Bibliography}

\begin{thebibliography}{10}

\bibitem{10049507}
Z.~Lai, S.~Li, X.~Tang, K.~Ge, W.~Liu, Y.~Duan, L.~Qiao, and D.~Li, ``{Merak: An Efficient Distributed {DNN} Training Framework With Automated 3D Parallelism for Giant Foundation Models},'' {\em IEEE Transactions on Parallel and Distributed Systems}, vol.~34, no.~5, pp.~1466--1478, 2023.

\bibitem{LIU2024317}
S.~Liu and T.~Ju, ``{APapo}: An asynchronous parallel optimization method for {DNN} models,'' {\em Future Generation Computer Systems}, vol.~152, pp.~317--330, 2024.

\bibitem{kim2023bpipe}
T.~Kim, H.~Kim, G.-I. Yu, and B.-G. Chun, ``{Bpipe: Memory-balanced pipeline parallelism for training large language models},'' in {\em International Conference on Machine Learning}, pp.~16639--16653, PMLR, 2023.

\bibitem{chilimbi2014project}
T.~Chilimbi, Y.~Suzue, J.~Apacible, and K.~Kalyanaraman, ``Project {A}dam: Building an efficient and scalable deep learning training system,'' in {\em 11th USENIX symposium on operating systems design and implementation (OSDI 14)}, pp.~571--582, 2014.

\bibitem{unnikrishnan2021layerpipe}
N.~K. Unnikrishnan and K.~K. Parhi, ``{LayerPipe}: Accelerating deep neural network training by intra-layer and inter-layer gradient pipelining and multiprocessor scheduling,'' in {\em Proceedings of the 2021 IEEE/ACM International Conference On Computer Aided Design (ICCAD)}, pp.~1--8, IEEE, 2021.

\bibitem{brakel2024model}
F.~Brakel, U.~Odyurt, and A.-L. Varbanescu, ``Model parallelism on distributed infrastructure: A literature review from theory to {LLM} case-studies,'' {\em arXiv preprint arXiv:2403.03699}, 2024.

\bibitem{shoeybi2019megatron}
M.~Shoeybi, M.~Patwary, R.~Puri, P.~LeGresley, J.~Casper, and B.~Catanzaro, ``Megatron-{LM}: Training multi-billion parameter language models using model parallelism,'' {\em arXiv preprint arXiv:1909.08053}, 2019.

\bibitem{yi2022optimizing}
X.~Yi, S.~Zhang, L.~Diao, C.~Wu, Z.~Zheng, S.~Fan, S.~Wang, J.~Yang, and W.~Lin, ``Optimizing {DNN} compilation for distributed training with joint op and tensor fusion,'' {\em IEEE Transactions on Parallel and Distributed Systems}, vol.~33, no.~12, pp.~4694--4706, 2022.

\bibitem{jiang2024megascale}
Z.~Jiang, H.~Lin, Y.~Zhong, Q.~Huang, Y.~Chen, Z.~Zhang, Y.~Peng, X.~Li, C.~Xie, S.~Nong, {\em et~al.}, ``{MegaScale}: Scaling large language model training to more than 10,000 {GPUs},'' in {\em Proceedings of the 21st USENIX Symposium on Networked Systems Design and Implementation (NSDI 24)}, pp.~745--760, 2024.

\bibitem{narayanan2019pipedream}
D.~Narayanan, A.~Harlap, A.~Phanishayee, V.~Seshadri, N.~R. Devanur, G.~R. Ganger, P.~B. Gibbons, and M.~Zaharia, ``{PipeDream}: generalized pipeline parallelism for {DNN} training,'' in {\em Proceedings of the 27th ACM Symposium on Operating Systems Principles}, SOSP'19, (New York, NY, USA), p.~1–15, 2019.

\bibitem{huang2019gpipe}
Y.~Huang, Y.~Cheng, A.~Bapna, O.~Firat, D.~Chen, M.~Chen, H.~Lee, J.~Ngiam, Q.~V. Le, Y.~Wu, {\em et~al.}, ``G{P}ipe: Efficient training of giant neural networks using pipeline parallelism,'' {\em Advances in Neural Information Processing Systems}, vol.~32, 2019.

\bibitem{boral2023anomaly}
S.~Boral, S.~Poddar, and A.~Ghosh, ``Anomaly detection in streaming environment by evolving neural network with interim decision,'' in {\em 2023 IEEE Region 10 Symposium (TENSYMP)}, pp.~1--6, IEEE, 2023.

\bibitem{zhao2021v}
S.~Zhao, F.~Li, X.~Chen, X.~Guan, J.~Jiang, D.~Huang, Y.~Qing, S.~Wang, P.~Wang, G.~Zhang, {\em et~al.}, ``{v Pipe: A virtualized acceleration system for achieving efficient and scalable pipeline parallel {DNN} training},'' {\em IEEE Transactions on Parallel and Distributed Systems}, vol.~33, no.~3, pp.~489--506, 2021.

\bibitem{ben2019demystifying}
T.~Ben-Nun and T.~Hoefler, ``Demystifying parallel and distributed deep learning: An in-depth concurrency analysis,'' {\em ACM Computing Surveys (CSUR)}, vol.~52, no.~4, pp.~1--43, 2019.

\bibitem{chen2022sapipe}
Y.~Chen, C.~Xie, M.~Ma, J.~Gu, Y.~Peng, H.~Lin, C.~Wu, and Y.~Zhu, ``{SAPipe: Staleness-aware pipeline for data parallel DNN training},'' {\em Advances in neural information processing systems}, vol.~35, pp.~17981--17993, 2022.

\bibitem{LI2021206}
Z.~Li, V.~Chang, H.~Hu, M.~Fu, J.~Ge, and F.~Piccialli, ``Optimizing makespan and resource utilization for multi-{DNN} training in gpu cluster,'' {\em Future Generation Computer Systems}, vol.~125, pp.~206--220, 2021.

\bibitem{raina2009large}
R.~Raina, A.~Madhavan, and A.~Y. Ng, ``Large-scale deep unsupervised learning using graphics processors,'' in {\em Proceedings of the 26th annual international conference on machine learning}, pp.~873--880, 2009.

\bibitem{jacobs2024system}
S.~A. Jacobs, M.~Tanaka, C.~Zhang, M.~Zhang, R.~Y. Aminadabi, S.~L. Song, S.~Rajbhandari, and Y.~He, ``{System optimizations for enabling training of extreme long sequence transformer models},'' in {\em Proceedings of the 43rd ACM Symposium on Principles of Distributed Computing}, pp.~121--130, 2024.

\bibitem{shallue2019measuring}
C.~J. Shallue, J.~Lee, J.~Antognini, J.~Sohl-Dickstein, R.~Frostig, and G.~E. Dahl, ``Measuring the effects of data parallelism on neural network training,'' {\em Journal of Machine Learning Research}, vol.~20, no.~112, pp.~1--49, 2019.

\bibitem{li2020pytorch}
S.~Li, Y.~Zhao, R.~Varma, O.~Salpekar, P.~Noordhuis, T.~Li, A.~Paszke, J.~Smith, B.~Vaughan, P.~Damania, {\em et~al.}, ``Pytorch distributed: Experiences on accelerating data parallel training,'' {\em arXiv preprint arXiv:2006.15704}, 2020.

\bibitem{cui2016geeps}
H.~Cui, H.~Zhang, G.~R. Ganger, P.~B. Gibbons, and E.~P. Xing, ``Gee{PS}: Scalable deep learning on distributed gpus with a gpu-specialized parameter server,'' in {\em Proceedings of the eleventh european conference on computer systems}, pp.~1--16, 2016.

\bibitem{dean2012large}
J.~Dean, G.~Corrado, R.~Monga, K.~Chen, M.~Devin, M.~Mao, M.~Ranzato, A.~Senior, P.~Tucker, K.~Yang, {\em et~al.}, ``Large scale distributed deep networks,'' {\em Advances in Neural Information Processing Systems}, vol.~25, 2012.

\bibitem{li2013parameter}
M.~Li, L.~Zhou, Z.~Yang, A.~Li, F.~Xia, D.~G. Andersen, and A.~Smola, ``Parameter server for distributed machine learning,'' in {\em Big learning NIPS workshop}, vol.~6, Lake Tahoe, CA, 2013.

\bibitem{paszke2019pytorch}
A.~Paszke, S.~Gross, F.~Massa, A.~Lerer, J.~Bradbury, G.~Chanan, T.~Killeen, Z.~Lin, N.~Gimelshein, L.~Antiga, {\em et~al.}, ``Pytorch: An imperative style, high-performance deep learning library,'' {\em Advances in neural information processing systems}, vol.~32, 2019.

\bibitem{li2023colossal}
S.~Li, H.~Liu, Z.~Bian, J.~Fang, H.~Huang, Y.~Liu, B.~Wang, and Y.~You, ``Colossal-{AI}: A unified deep learning system for large-scale parallel training,'' in {\em Proceedings of the 52nd International Conference on Parallel Processing}, pp.~766--775, 2023.

\bibitem{dutta2024timeprest}
A.~Dutta, N.~Chaki, and R.~K. De, ``Ti{M}e{PR}e{S}t: Time and memory efficient pipeline parallel {DNN} training with removed staleness,'' {\em arXiv preprint arXiv:2410.14312}, 2024.
\newblock Submitted to FGCS (Under Review).

\bibitem{AKINTOYE2023432}
S.~Akintoye, L.~Han, H.~Lloyd, X.~Zhang, D.~Dancey, H.~Chen, and D.~Zhang, ``Layer-wise partitioning and merging for efficient and scalable deep learning,'' {\em Future Generation Computer Systems}, vol.~149, pp.~432--444, 2023.

\bibitem{chen2016revisiting}
J.~Chen, X.~Pan, R.~Monga, S.~Bengio, and R.~Jozefowicz, ``Revisiting distributed synchronous sgd,'' {\em arXiv preprint arXiv:1604.00981}, 2016.

\bibitem{wan2025pipeoffload}
X.~Wan, P.~Qi, G.~Huang, J.~Li, and M.~Lin, ``{PipeOffload: Improving Scalability of Pipeline Parallelism with Memory Optimization},'' {\em arXiv preprint arXiv:2503.01328}, 2025.

\bibitem{narayanan2021memory}
D.~Narayanan, A.~Phanishayee, K.~Shi, X.~Chen, and M.~Zaharia, ``Memory-efficient pipeline-parallel {DNN} training,'' in {\em Proceedings of the International Conference on Machine Learning}, pp.~7937--7947, PMLR, 2021.

\bibitem{10.1145/3472456.3472497}
X.~Ye, Z.~Lai, S.~Li, L.~Cai, D.~Sun, L.~Qiao, and D.~Li, ``{Hippie: A Data-Paralleled Pipeline Approach to Improve Memory-Efficiency and Scalability for Large {DNN} Training},'' in {\em Proceedings of the 50th International Conference on Parallel Processing}, ICPP '21, (New York, NY, USA), Association for Computing Machinery, 2021.

\bibitem{zhang2023pipepar}
J.~Zhang, G.~Niu, Q.~Dai, H.~Li, Z.~Wu, F.~Dong, and Z.~Wu, ``{PipePar}: Enabling fast {DNN} pipeline parallel training in heterogeneous gpu clusters,'' {\em Neurocomputing}, vol.~555, p.~126661, 2023.

\bibitem{tangkoala}
Y.~Tang, L.~Yin, Q.~Li, H.~Zhu, H.~Li, X.~Zhang, L.~Qiao, D.~Li, and J.~Li, ``{Koala: Efficient Pipeline Training through Automated Schedule Searching on Domain-Specific Language},'' {\em ACM Transactions on Architecture and Code Optimization}.

\bibitem{liu2025mario}
W.~Liu, M.~Li, G.~Tan, and W.~Jia, ``{Mario: Near Zero-cost Activation Checkpointing in Pipeline Parallelism},'' in {\em Proceedings of the 30th ACM SIGPLAN Annual Symposium on Principles and Practice of Parallel Programming}, pp.~197--211, 2025.

\bibitem{lin2025weipipe}
J.~Lin, Z.~Liu, Y.~You, J.~Wang, W.~Zhang, and R.~Zhao, ``{WeiPipe: Weight Pipeline Parallelism for Communication-Effective Long-Context Large Model Training},'' in {\em Proceedings of the 30th ACM SIGPLAN Annual Symposium on Principles and Practice of Parallel Programming}, pp.~225--238, 2025.

\bibitem{liu2024deepseek}
A.~Liu, B.~Feng, B.~Xue, B.~Wang, B.~Wu, C.~Lu, C.~Zhao, C.~Deng, C.~Zhang, C.~Ruan, {\em et~al.}, ``Deepseek-v3 technical report,'' {\em arXiv preprint arXiv:2412.19437}, 2024.

\bibitem{sun2025mepipe}
Z.~Sun, S.~Chen, Y.~Wang, J.~Sha, G.~Feng, and W.~Chen, ``Mepipe: Democratizing llm training with memory-efficient slice-level pipeline scheduling on cost-effective accelerators,'' in {\em Proceedings of the Twentieth European Conference on Computer Systems}, pp.~1263--1278, 2025.

\bibitem{qi2024zero}
P.~Qi, X.~Wan, G.~Huang, and M.~Lin, ``Zero bubble (almost) pipeline parallelism,'' in {\em The Twelfth International Conference on Learning Representations}, 2024.

\bibitem{jeon2025graphpipe}
B.~Jeon, M.~Wu, S.~Cao, S.~Kim, S.~Park, N.~Aggarwal, C.~Unger, D.~Arfeen, P.~Liao, X.~Miao, {\em et~al.}, ``Graph{P}ipe: Improving performance and scalability of dnn training with graph pipeline parallelism,'' in {\em Proceedings of the 30th ACM International Conference on Architectural Support for Programming Languages and Operating Systems, Volume 1}, pp.~557--571, 2025.

\bibitem{lin2025enhancing}
X.~Lin, C.~Li, Z.~Huang, C.~Wang, B.~Xiao, H.~Yang, S.~Duan, and Y.~Liu, ``{Enhancing Memory Efficiency in Large Language Model Training Through Chronos-aware Pipeline Parallelism},'' {\em arXiv preprint arXiv:2503.03182}, 2025.

\end{thebibliography}

\appendix

\section{Appendix}
\label{sec:Appendix}
\subsection{Technical Proof}
\label{sec:Technical Proof}
\subsubsection{Mathematical expression for computing the significance of stale weights}
\label{sec:Mathematical expression for computing the significance of stale weights}
Let $\delta$ be the degree of staleness and $f(\delta)$ be the influence of the degree of staleness $\delta$. We assume $\delta \geq 0$ and $f(\delta) \in (0,1]$. At any given point, the rate of decay is determined by how much remain. For example, if we have more remaining quantity (large $f(\delta))$, the decay is faster. As the quantity decreases, the decay slows down. In the context of gradient staleness mitigation, where older information gradually looses its relevance, we can assume that the amount of influence an older gradient should have currently, that depends on how much influence it had at earlier time-points and how much is remaining. The absolute value of the rate of change of the influence $\Big|\dfrac{f(\delta + \Delta\delta) -f(\delta)}{\Delta\delta}\Big|$, is proportional to the current value of $f(\delta)$. The assumption can be expressed mathematically as follows.
\begin{equation}\label{eqn:initial assumption}
    \Big|\dfrac{f(\delta + \Delta\delta) -f(\delta)}{\Delta\delta}\Big| \propto f(\delta)
\end{equation}
This implies
\begin{equation}\label{eqn:assumption lambda}
     \Big|\dfrac{f(\delta + \Delta\delta) -f(\delta)}{\Delta\delta}\Big| = \pm\lambda \cdot f(\delta)
\end{equation}
where $\lambda > 0$ is a proportionality constant. 
Equation \ref{eqn:assumption lambda} can be re-written as
\begin{equation}\label{eqn:difference equation representation}
\begin{split}
    f(\delta + \Delta\delta) -f(\delta) = \pm\lambda \cdot {\Delta\delta} \cdot f(\delta) \\
    f(\delta + \Delta\delta) = f(\delta)\Big(1 \pm\lambda \cdot {\Delta\delta}\Big) \\
\end{split}
\end{equation}
For $\delta=0$, $f(\delta) = 1$ since $\delta=0$ means no staleness and $f(\delta)$ indicates the significance of the weights which are stale of degree $\delta$. Thus, $f(0)$ indicates the significance of the current weights. From Equation \ref{eqn:difference equation representation} we can write
\begin{equation}\label{eqn:f in terms of f(0)}
    \begin{split}
        f(\Delta\delta) = f(0)\Big(1 \pm\lambda \cdot {\Delta\delta}\Big) \\
        f(1+\Delta\delta) = f(1)\Big(1 \pm\lambda \cdot {\Delta\delta}\Big)
            &= f(0)\Big(1 \pm\lambda \cdot {\Delta\delta}\Big)^2 \\
            & \vdots \\
            f(\delta - 1 + \Delta\delta) = f(0)\Big(1 \pm\lambda \cdot {\Delta\delta}\Big)^\delta
                      &= \Big(1 \pm\lambda \cdot {\Delta\delta}\Big)^\delta
    \end{split}
\end{equation}
Let, 
\begin{equation}\label{eqn:expression for delta}
    \begin{split}
        \delta = n \cdot \Delta\delta,
    \end{split}
\end{equation}
where $\delta$ is the total amount of staleness, and $\Delta\delta$ is the step-size.
Equation \ref{eqn:f in terms of f(0)} can be rewritten as
\begin{equation}\label{eqn:updated f()}
    \begin{split}
        f(\delta - 1 + \Delta\delta) = \Big(1 \pm\lambda \cdot \frac{\delta}{n}\Big)^{n \cdot \Delta\delta}
    \end{split}
\end{equation}
Considering $\Delta\delta = 1$, Equation \ref{eqn:updated f()} can be re-written as
\begin{equation}\label{eqn:final f(delta)}
\begin{split}
    f(\delta) = \Big(1 \pm\lambda \cdot \frac{\delta}{n}\Big)^n \\
    \lim_{n\to\infty} f(\delta) = \lim_{n\to\infty} \Big(1 + \frac{(\pm\lambda \cdot \delta)}{n}\Big)^n = e^{\pm\lambda\delta} 
\end{split}
\end{equation}
Thus for large n, we can write
\begin{equation}\label{eqn:possible expressions for f(delta)}
    f(\delta) \approx e^{\pm\lambda\delta} 
\end{equation}
If $f(\delta) \approx e^{\lambda\delta}$ and $\delta \rightarrow \infty$, then $f(\delta) \rightarrow \infty$.
Since we assume $f(\delta) \in (0,1]$, thus, it is not a feasible case.
Thus, the final expression is 
\begin{equation}\label{eqn:final expression of f(delta)}
    f(\delta) \approx e^{-\lambda\delta} 
\end{equation}
\subsubsection{Mathematical expression for the intermediate weights}
\label{sec:Mathematical expression for the intermediate weights}
For $\delta \rightarrow 0$, $f(0) \rightarrow 1$. It indicates that there is no staleness. If $\delta \rightarrow \infty$, $f(0) \rightarrow 0$. It indicates that very stale gradients have negligible influence. We intend to remove the unnecessary contribution of the stale weight during weight updation. Thus, we scale the stale weight by its significance or influence, and the rest, i.e, by $(1-f(\delta))$,
\vspace{0.5em}
\begin{equation}\label{eqn:final expression of W(x,y)}
\resizebox{0.43\textwidth}{!}{$\begin{split}
   \textbf{W}_{i}(x,y) &= f(\delta)\times\textbf{W}_{i}(x|y)+(1-f(\delta))\times(\textbf{W}_{i}(x,y)-\textbf{W}_{i}(x|y)) \\
    &= (2 - \frac{1}{f(\delta)})\times\textbf{W}_{i}(x|y)
\end{split}$}
\end{equation}
\vspace{0.5em}
Here the term $\textbf{W}_{i}(x|y)$, for a worker machine $i$, denotes the updated weights computed using mini-batch $x$, which becomes stale once the updates using mini-batch $y$ is available.
The term $\textbf{W}_{i}(x,y)$ stands for the intermediate weights between the stale weights $\textbf{W}_{i}(x|y)$ and the latest updated weights. Here $\textbf{W}_{i}(x,y)$ is computed based on $\textbf{W}_{i}(x|y)$ and the degree of staleness $\delta \geq 0$. The backward pass on a mini-batch is performed based on $\textbf{W}_{i}(x,y)$. We assume that the $\textbf{W}_{i}(x,y)$ is a combination of the stale weights $\textbf{W}_{i}(x|y)$ and other potential factors influencing the training process.
\\As we know,
\begin{equation}\label{eqn:range of f(delta)}
    \begin{split}
        0<f(\delta)\leq 1 \\
        \infty>\frac{1}{f(\delta)}\geq 1 \\
        -\infty<-\frac{1}{f(\delta)}\leq -1 \\
        -\infty<2-\frac{1}{f(\delta)}\leq 1
    \end{split}
\end{equation}

\subsection{Supplementary Experimental Results}
\label{sec:Supplementary Experimental Results}
\begin{figure}[h]
  \centering
  \begin{subfigure}{.25\textwidth} 
    \centering
    \includegraphics[width=\linewidth, height=7cm]{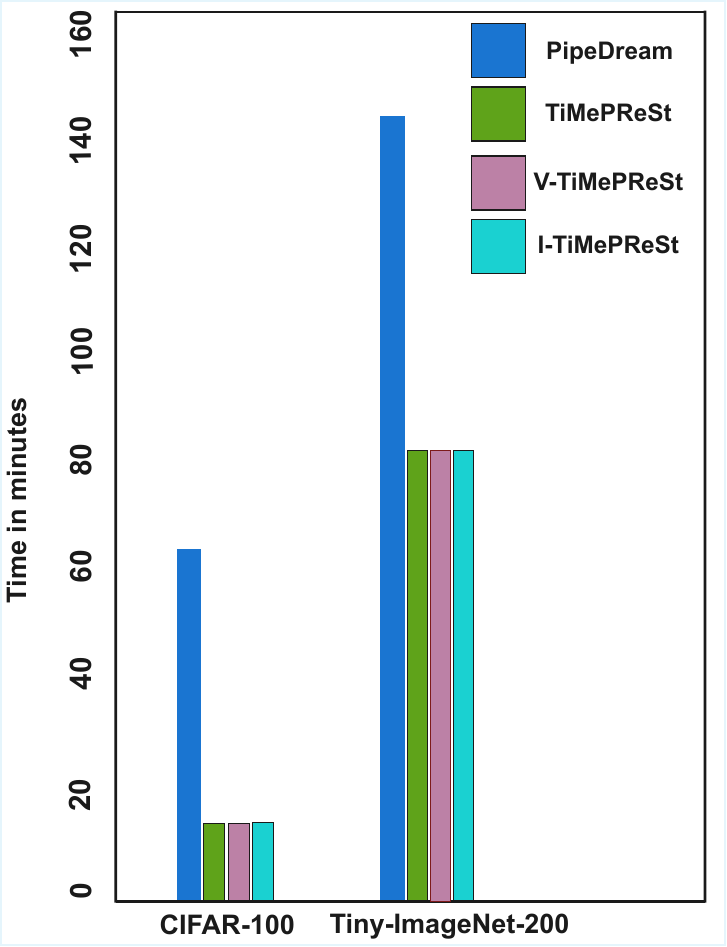}
    \caption{Hardware Efficiency (Time per epoch)}
    \label{fig:Time per epoch ResNet-50}
  \end{subfigure}%
  \hfill
  \begin{subfigure}{.22\textwidth}
    \centering
    \includegraphics[width=\linewidth, height=5.5cm]{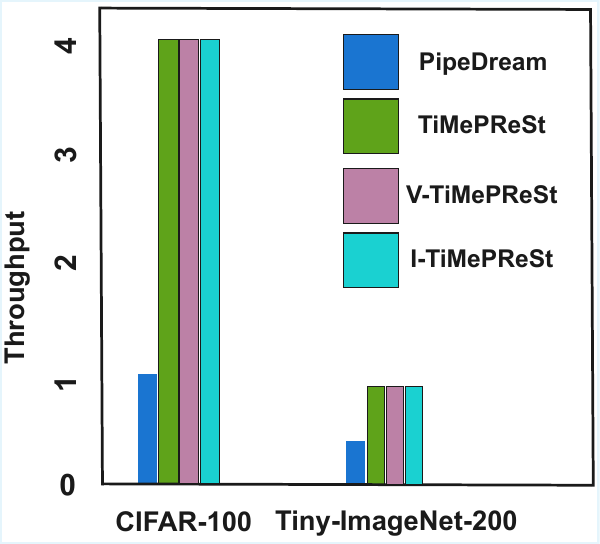}
    \caption{Throughput (epochs/hr)}
    \label{fig:Throughput ResNet-50}
  \end{subfigure}
  \caption{\textbf{Comparison among V-TiMePReSt, I-TiMePReSt, TiMePReSt and PipeDream platforms based on hardware efficiency and throughput for training ResNet-50 on CIFAR-100 and Tiny-ImageNet-200 datasets.}}
  \label{fig:Comparison between V-TiMePReSt, I-TiMePReSt, TiMePReSt and PipeDream (ResNet-50 on  CIFAR-100 and Tiny-ImageNet-200 datasets), based on hardware efficiency and throughput}
\end{figure}
\begin{figure}[h]
  \centering
   \includegraphics[width=1.0\linewidth]{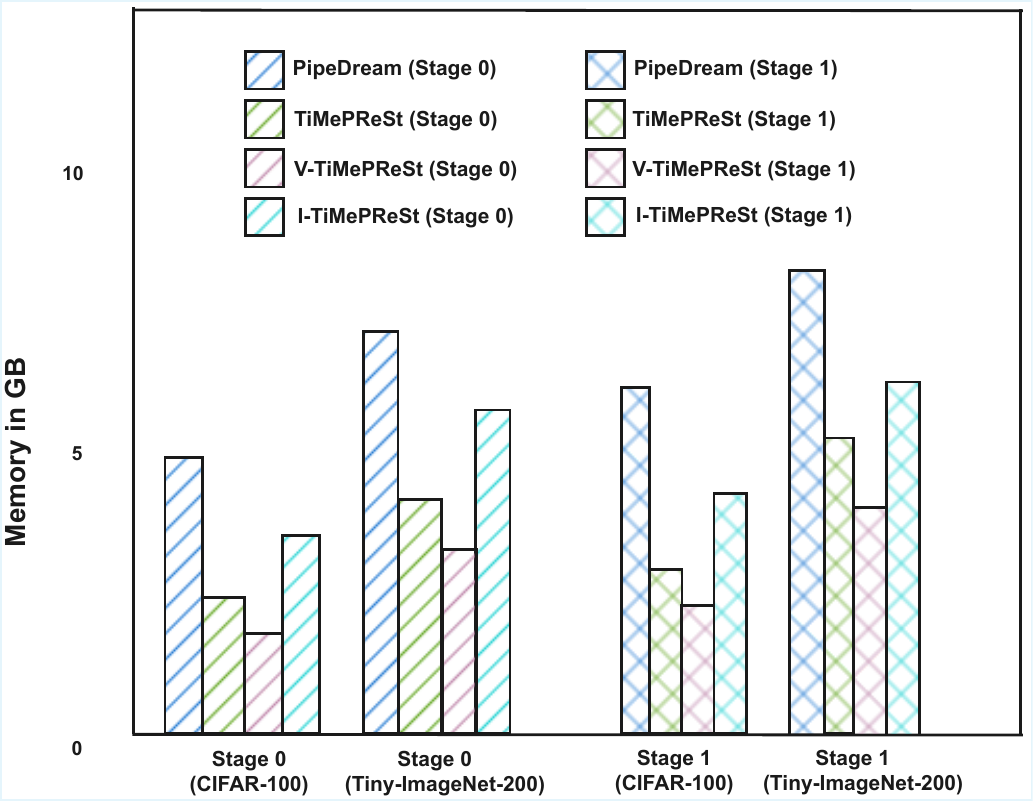}
  \caption{\textbf{Memory footprint per stage using two worker nodes for training ResNet-50 on CIFAR-100 and Tiny-ImageNet-200 image classification datasets, for V-TiMePReSt, I-TiMePReSt, TiMePReSt and PipeDream platforms.}}
  \label{fig:Memory consumption comparison for ResNet-50}
\end{figure}
\begin{figure}[h]
  \begin{subfigure}{.23\textwidth}
  \centering
    \includegraphics[width=\linewidth]{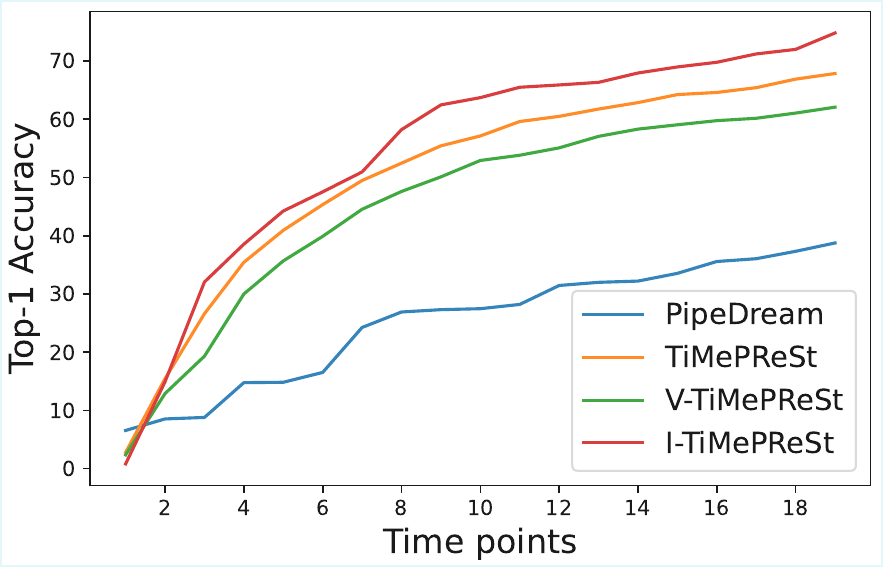}
    \caption{Top-1 accuracy vs time}
    \label{fig:Top-1 accuracy to time_cifar100_ResNet50}
  \end{subfigure}
  \begin{subfigure}{.23\textwidth}
  \centering
    \includegraphics[width=\linewidth]{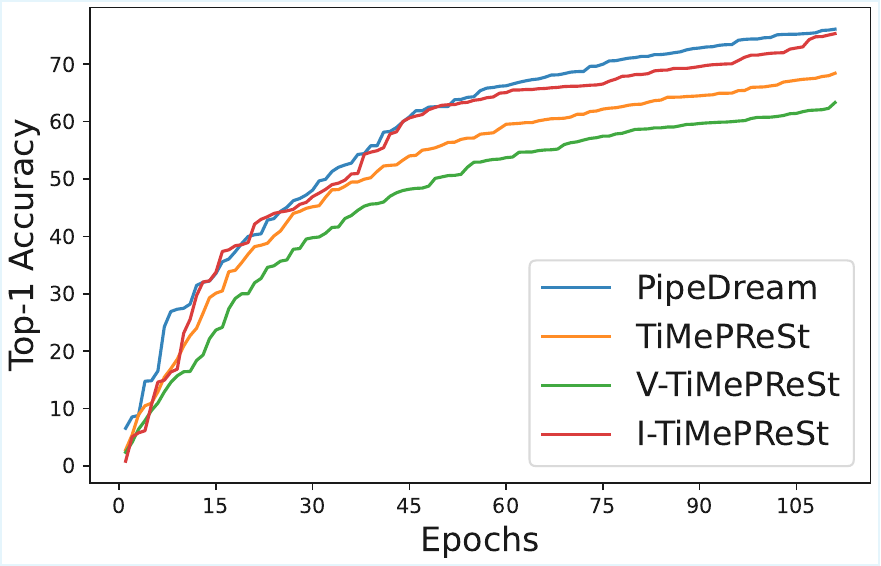}
    \caption{Top-1 accuracy vs epoch}
    \label{fig:Top-1 accuracy to epoch_cifar100_ResNet50}
  \end{subfigure}
  \begin{subfigure}{.23\textwidth}
  \centering
    \includegraphics[width=\linewidth]{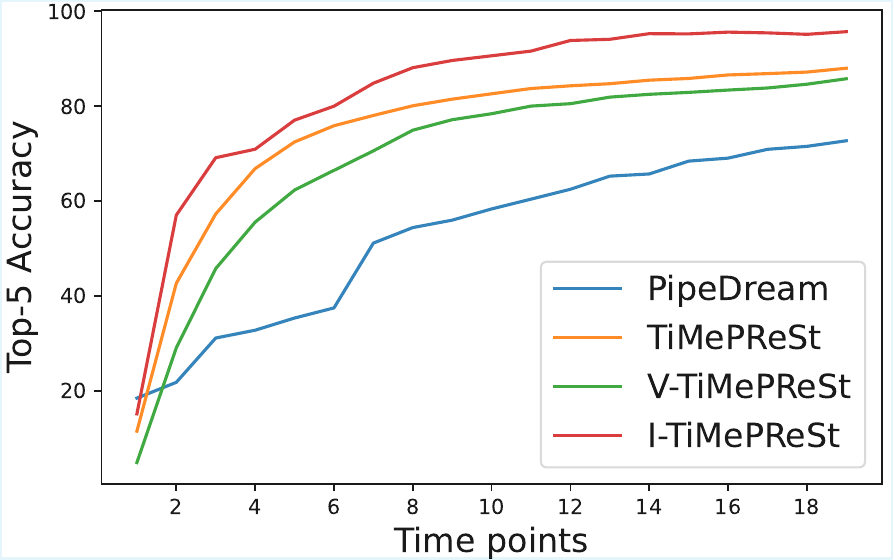}
    \caption{Top-5 accuracy vs time}
    \label{fig:Top-5 accuracy to time_cifar100_ResNet50}
  \end{subfigure}
  \begin{subfigure}{.23\textwidth}
  \centering
    \includegraphics[width=\linewidth]{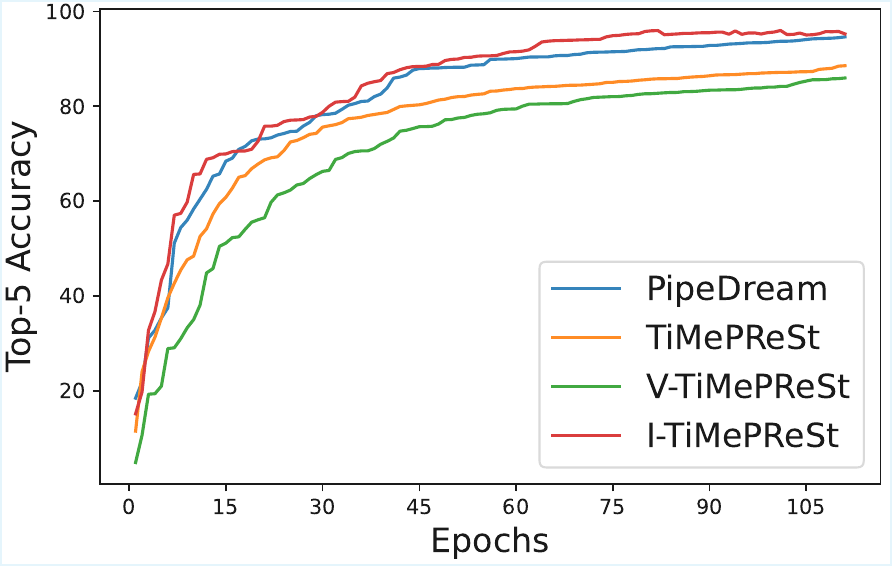}
    \caption{Top-5 accuracy vs epoch}
    \label{fig:Top-5 accuracy to epoch_cifar100_ResNet50}
  \end{subfigure}
  \begin{subfigure}{.23\textwidth}
  \centering
    \includegraphics[width=\linewidth]{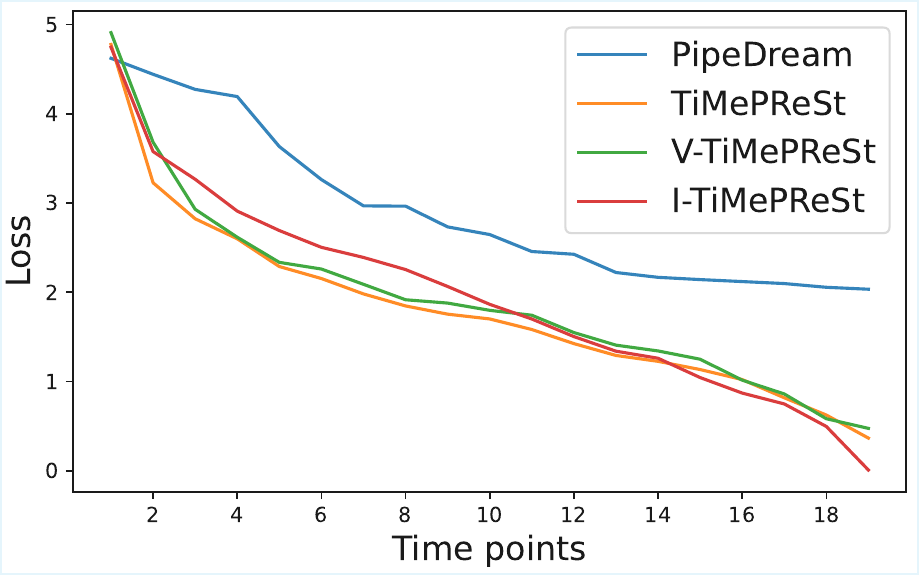}
    \caption{Loss vs time}
    \label{fig:Loss comparison_time_cifar100_ResNet50}
  \end{subfigure}
  \hspace{0.7em}
  \begin{subfigure}{.23\textwidth}
  \centering
    \includegraphics[width=\linewidth]{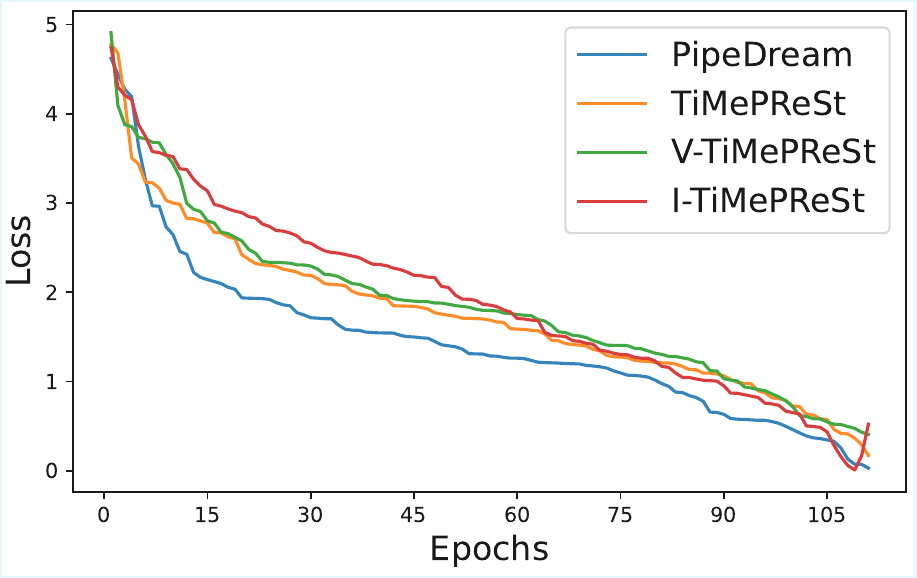}
    \caption{Loss vs epoch}
    \label{fig:Loss comparison_epoch_cifar100_ResNet50}
  \end{subfigure}
  \caption{\textbf{Performance comparison of V-TiMePReSt, I-TiMePReSt, TiMePReSt, and PipeDream platforms for ResNet-50 training on CIFAR-100 dataset based on accuracy and loss achieved over time and epochs. Figures \ref{fig:Top-1 accuracy to time_cifar100_ResNet50}, \ref{fig:Top-5 accuracy to time_cifar100_ResNet50} and \ref{fig:Loss comparison_time_cifar100_ResNet50} represent the plots with respect to time, whereas Figures \ref{fig:Top-1 accuracy to epoch_cifar100_ResNet50}, \ref{fig:Top-5 accuracy to epoch_cifar100_ResNet50} and \ref{fig:Loss comparison_epoch_cifar100_ResNet50} represent the plots with respect to epochs (statistical efficiency).}}
  \label{fig:Performance Comparison of V-TiMePReSt, I-TiMePReSt, TiMePReSt and PipeDream (ResNet-50 on CIFAR-100)}
\end{figure}
\begin{figure}[h]
  \begin{subfigure}{.23\textwidth}
  \centering
    \includegraphics[width=\linewidth]{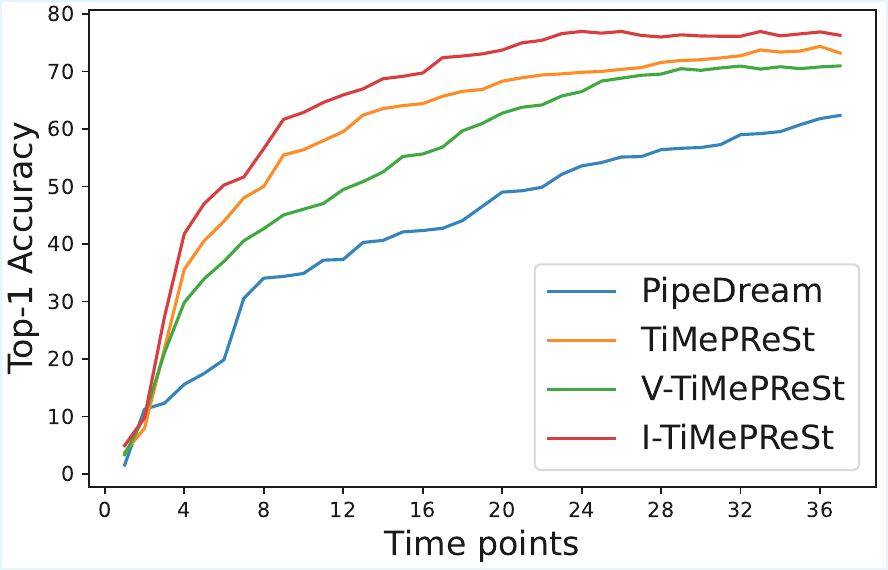}
    \caption{Top-1 accuracy vs time}
    \label{fig:Top-1 accuracy to time_tiny_ImageNet_ResNet50}
  \end{subfigure}
  \begin{subfigure}{.23\textwidth}
  \centering
    \includegraphics[width=\linewidth]{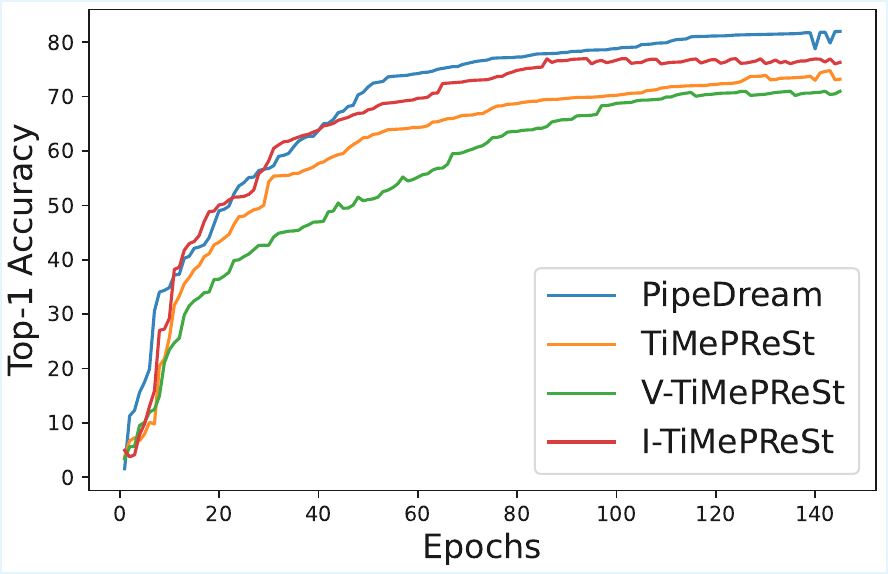}
    \caption{Top-1 accuracy vs epoch}
    \label{fig:Top-1 accuracy to epoch_TinyImageNet200_ResNet50}
  \end{subfigure}
  \begin{subfigure}{.23\textwidth}
  \centering
    \includegraphics[width=\linewidth, height=2.7cm]{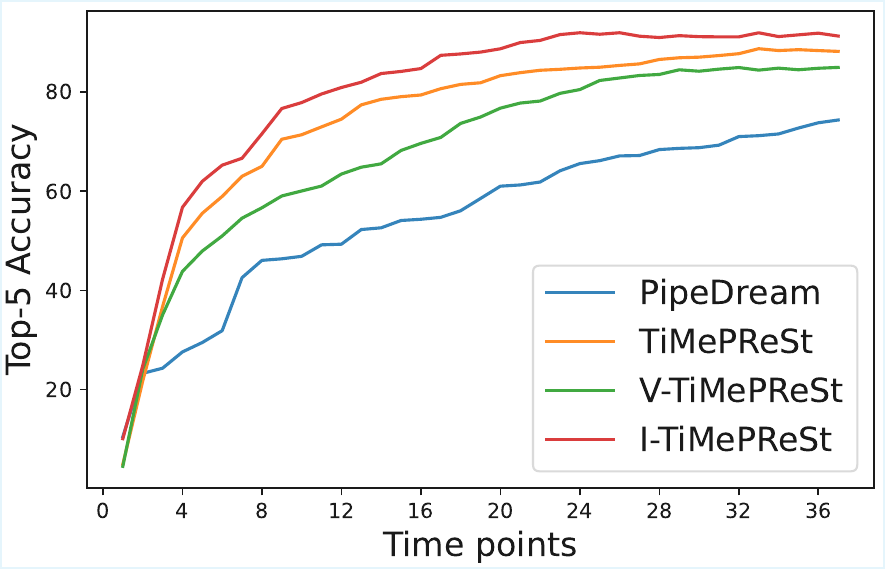}
    \caption{Top-5 accuracy vs time}
    \label{fig:Top-5 accuracy to time_tiny_ImageNet_ResNet50}
  \end{subfigure}
  \begin{subfigure}{.23\textwidth}
  \centering
    \includegraphics[width=\linewidth]{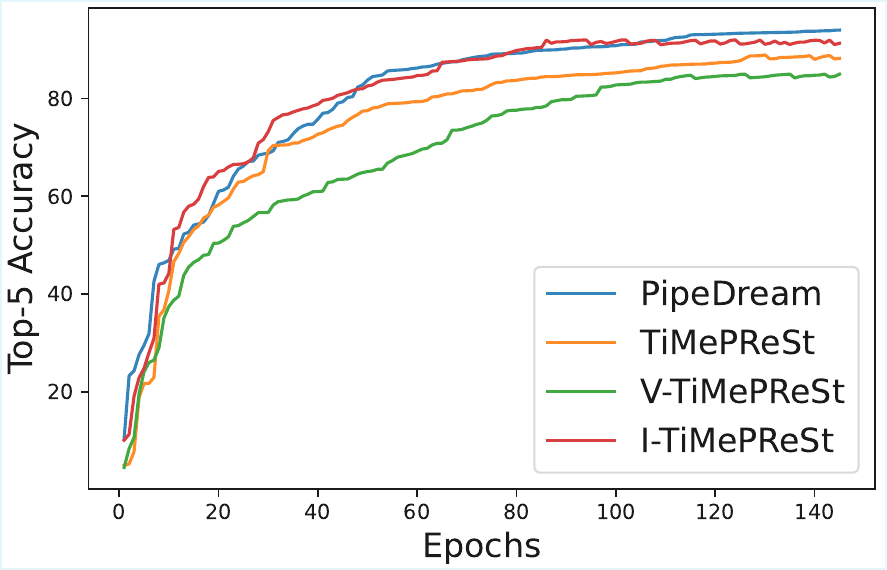}
    \caption{Top-5 accuracy vs epoch}
    \label{fig:Top-5 accuracy to epoch_TinyImageNet200_ResNet50}
  \end{subfigure}
  \begin{subfigure}{.23\textwidth}
  \centering
    \includegraphics[width=\linewidth]{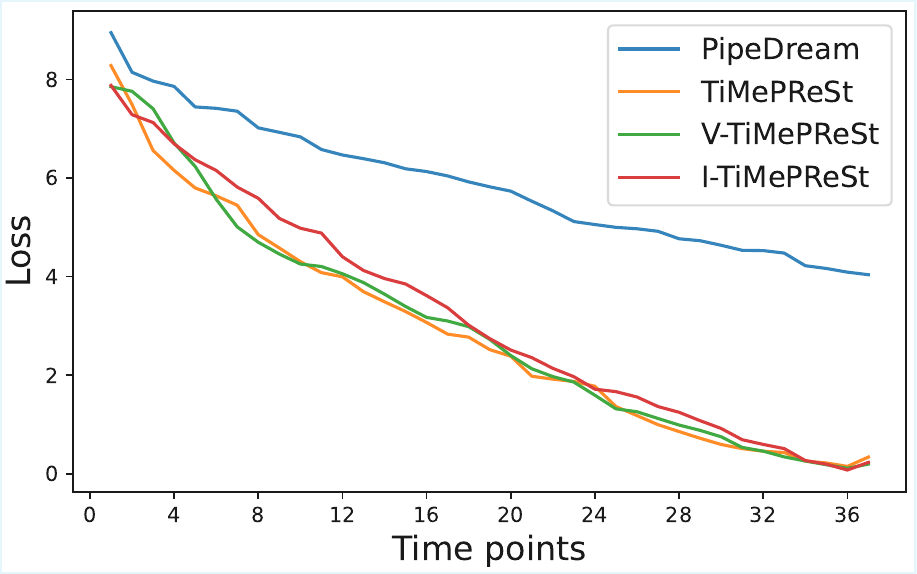}
    \caption{Loss vs time}
    \label{fig:Loss comparison_time_tiny_ImageNet_ResNet50}
  \end{subfigure}
  \hspace{0.6em}
  \begin{subfigure}{.23\textwidth}
  \centering
    \includegraphics[width=\linewidth]{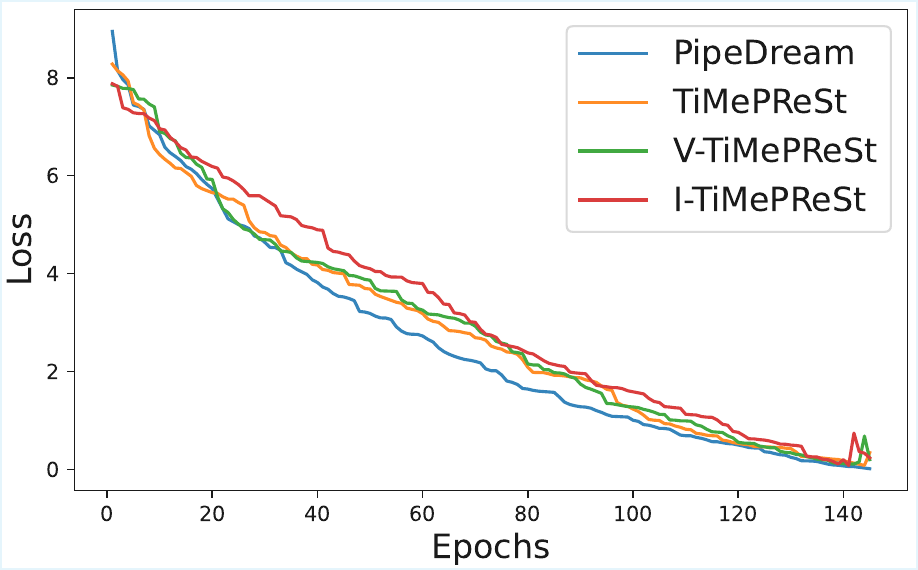}
    \caption{Loss vs epoch}
    \label{fig:Loss comparison_epoch_TinyImageNet200_ResNet50}
  \end{subfigure}
  \caption{\textbf{Performance comparison of V-TiMePReSt, I-TiMePReSt, TiMePReSt, and PipeDream platforms for ResNet-50 on Tiny-ImageNet-200 datasets based on accuracy and loss achieved over time and epochs. Figures \ref{fig:Top-1 accuracy to time_tiny_ImageNet_ResNet50}, \ref{fig:Top-5 accuracy to time_tiny_ImageNet_ResNet50} and \ref{fig:Loss comparison_time_tiny_ImageNet_ResNet50} represent the plots with respect to time, whereas Figures \ref{fig:Top-1 accuracy to epoch_TinyImageNet200_ResNet50}, \ref{fig:Top-5 accuracy to epoch_TinyImageNet200_ResNet50} and \ref{fig:Loss comparison_epoch_TinyImageNet200_ResNet50} represent the plots with respect to epochs (statistical efficiency).}}
  \label{fig:Performance Comparison of V-TiMePReSt, I-TiMePReSt, TiMePReSt and PipeDream (ResNet50 on Tiny_ImageNet)}
\end{figure}


\end{document}